\documentclass[floats,floatfix,showpacs,amssymb,prd,twocolumn,superscriptaddress,nofootinbib,nolongbibliography,reprint]{revtex4-1}

\usepackage{amssymb,amsmath,verbatim,mathtools,needspace,enumitem,etoolbox,graphicx,physics,microtype,afterpage,xspace,tabularx,lmodern,multirow}
\usepackage{gensymb}
\usepackage[dvipsnames, usenames]{xcolor}
\definecolor{linkcolor}{rgb}{0.0,0.3,0.5}
\usepackage[unicode, colorlinks=true, linkcolor=linkcolor, citecolor=linkcolor, filecolor=linkcolor, urlcolor=linkcolor, linktocpage, breaklinks]{hyperref}
\usepackage[all]{hypcap}
\usepackage[T1]{fontenc}
\usepackage[utf8]{inputenc}
\usepackage[usenames,dvipsnames]{xcolor}
\hypersetup{colorlinks=true,citecolor=romared,linkcolor=romared,urlcolor=romared}

\definecolor{romared}{RGB}{142,0,28}

\newcommand{\be}{\begin{equation}}
\newcommand{\ee}{\end{equation}}

\def\be{\begin{equation}}
\def\ee{\end{equation}}
\newcommand{\beq}{\begin{eqnarray}}
\newcommand{\eeq}{\end{eqnarray}}

\usepackage{aas_macros}
\usepackage{makecell}
\usepackage{soul}
\usepackage{comment}
\usepackage[nolist,nohyperlinks]{acronym}

\allowdisplaybreaks

\begin{document}

\pagenumbering{arabic}

\title{Analysis of Ringdown Overtones in GW150914}

\author{Roberto Cotesta}
\affiliation{Department of Physics and Astronomy, Johns Hopkins University,
3400 N. Charles Street, Baltimore, Maryland, 21218, USA}
\author{Gregorio Carullo}
\affiliation{Dipartimento di Fisica ``Enrico Fermi'', Universit\`a di Pisa, Pisa I-56127, Italy}
\affiliation{INFN sezione di Pisa, Pisa I-56127, Italy}
\affiliation{${}^1$Theoretisch-Physikalisches Institut, Friedrich-Schiller-Universit{\"a}t Jena, Fr{\"o}belstieg 1, 07743 Jena, Germany}
\author{Emanuele Berti}
\affiliation{Department of Physics and Astronomy, Johns Hopkins University,
3400 N. Charles Street, Baltimore, Maryland, 21218, USA}
\author{Vitor Cardoso}
\affiliation{CENTRA, Departamento de F\'{\i}sica, Instituto Superior T\'ecnico -- IST, Universidade de Lisboa -- UL,
Avenida Rovisco Pais 1, 1049-001 Lisboa, Portugal}
\affiliation{Niels Bohr International Academy, Niels Bohr Institute, Blegdamsvej 17, 2100 Copenhagen, Denmark}

\date{\today}

\begin{abstract}
  We analyze GW150914 post-merger data to understand if ringdown overtone detection claims are robust. We find no evidence in favor of an overtone in the data after the waveform peak. Around the peak, the Bayes factor does not indicate the presence of an overtone, while the support for a nonzero amplitude is sensitive to changes in the starting time much smaller than the overtone damping time. This suggests that claims of an overtone detection are noise-dominated. We perform GW150914-like injections in neighboring segments of the real detector noise, and we show that noise can indeed induce artificial evidence for an overtone.
\end{abstract}

\maketitle


\acrodef{LSC}[LSC]{LIGO Scientific Collaboration}
\acrodef{BH}{black hole}
\acrodef{NS}{neutron star}
\acrodef{PN}{Post-Newtonian}
\acrodef{BBH}{binary black-hole}
\acrodef{BNS}{binary neutron-star}
\acrodef{NSBH}{neutron-star black-hole}
\acrodef{EOB}{effective-one-body}
\acrodef{NR}{numerical relativity}
\acrodef{GW}{gravitational wave}
\acrodef{PSD}{power spectral density}
\acrodef{aLIGO}{Advanced Laser interferometer Gravitational-Wave Observatory}
\acrodef{AZDHP}{aLIGO zero detuned high power density}
\acrodef{GR}{general relativity}
\acrodef{PE}{parameter estimation}
\acrodef{LAL}{LIGO algorithm library}
\acrodef{TPI}{tensor-product interpolant}
\acrodef{SVD}{singular value decomposition}
\acrodef{SNR}{signal-to-noise ratio}
\acrodef{ODE}{ordinary differential equation}
\acrodef{PDE}{partial differential equation}
\acrodef{ROM}{reduced order model}
\acrodef{QNM}{quasi-normal mode}
\acrodef{IMR}{inspiral-merger-ringdown}
\acrodef{LVK}{LIGO-Virgo-KAGRA}
\acrodef{SXS}{Simulating eXtreme Spacetimes}

\newcommand{\pyRing}{\texttt{pyRing}\,\,}
\newcommand{\ringdown}{\texttt{ringdown}\,\,}
\newcommand{\hlm}{h_{\ell m}}
\newcommand{\ylm}{{}_{-2}Y_{\ell m}}

\noindent {\bf \em Introduction.}
Since the first detection of \acp{GW} from a binary \ac{BH} merger, GW150914~\cite{LIGOScientific:2016aoc},
the \ac{LVK} Collaboration~\cite{LIGOScientific:2014pky, VIRGO:2014yos, KAGRA:2020tym} reported 90 events with a probability of astrophysical origin $p_{\rm astro} > 0.5$ during the first three observing runs~\cite{LIGOScientific:2018mvr,LIGOScientific:2020ibl,LIGOScientific:2021usb,LIGOScientific:2021djp}.
These \ac{GW} signals, combined with those detected by independent groups~\cite{Nitz:2018imz,Nitz:2020oeq,Nitz:2021uxj,Venumadhav:2019lyq,Zackay:2019btq}, have broadened our understanding of cosmology~\cite{LIGOScientific:2021aug}, the astrophysics of compact objects~\cite{LIGOScientific:2021psn}, matter at supranuclear densities~\cite{Chatziioannou:2020pqz}, and \ac{GR} in the strong-field regime~\cite{LIGOScientific:2020tif}. 

Among the numerous tests of \ac{GR} proposed over the years, \ac{BH} spectroscopy with the so-called ``ringdown''
relaxation phase following the merger presents unique opportunities to characterize the remnant as a Kerr \ac{BH}. 
In linearized \ac{GR}, the two \ac{GW} polarizations $h_{+,\,\times}$ can be decomposed as
$h_+ - i h_\times \equiv \sum_{\ell m} \hlm(t) \ylm(\iota,\,\phi)$, 
where the (spin-weighted) spherical harmonics $\ylm (\iota, \phi)$ depend on two angles that characterize the direction from the source to the observer. Each multipolar component is a superposition of damped exponentials known as \acp{QNM}:
\begin{equation}\label{eq:model}
\hlm(t) \equiv \sum_n A_{\ell m n} e^{i\left[-\omega_{\ell m n}(t - t_{\ell m n}^\mathrm{start}) + \phi_{\ell m n}\right]}e^{-(t - t_{\ell m n}^\mathrm{start})/\tau_{\ell m n}},
\end{equation}
where we ignored spherical-spheroidal mode-mixing between different corotating $\ell$ modes, and the contribution of counterrotating modes (a valid assumption for GW150914).
In \ac{GR}, the \ac{QNM} frequencies $\omega_{\ell m n}$ and damping times $\tau_{\ell m n}$ depend only on the remnant \ac{BH}'s mass $M_\mathrm{f}$ and spin $a_\mathrm{f}$~\cite{Press:1971wr,Chandrasekhar:1975zza,Detweiler:1980gk,Kokkotas:1999bd,Dreyer:2003bv,Berti:2005ys,Berti:2009kk}.
The QNM amplitudes $A_{\ell m n}$ and phases $\phi_{\ell m n}$ were unknown before the first numerical \ac{BH} merger simulations, and early work on BH spectroscopy~\cite{Berti:2005ys} had to rely on educated guesses~\cite{Flanagan:1997sx}. We now know that radiation from a binary \ac{BH} merger is dominated by the $\ell=|m|=2$ component, while higher multipoles are subdominant~\cite{Buonanno:2006ui,Berti:2007fi}. For fixed $(\ell,\,m)$, the \acp{QNM} are sorted by the magnitude of $\tau_{\ell m n}$: the fundamental mode ($n=0$) has the longest damping time, and the integer $n$ labels the so-called ``overtones.''

\begin{figure*}[thbp]
         \centering
         \includegraphics[width=\textwidth]{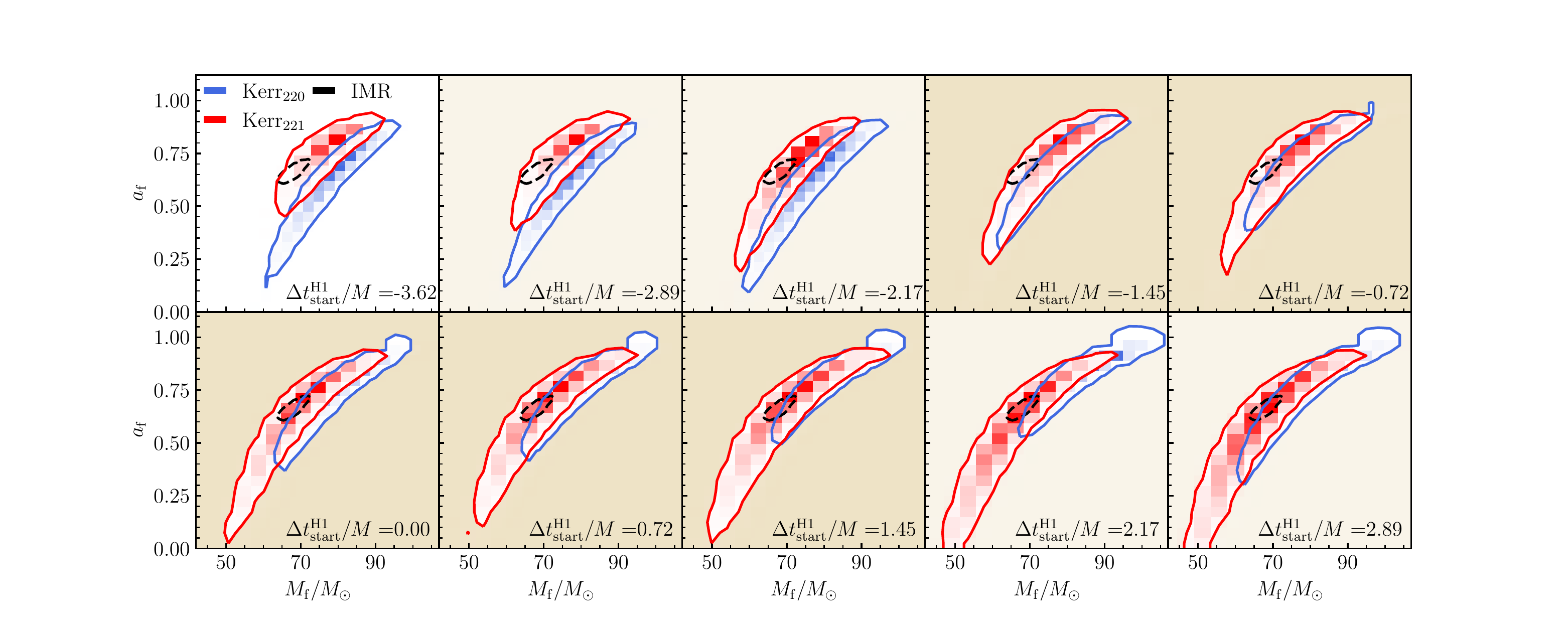}
   \caption{Mass and spin of the remnant \ac{BH} for GW150914. Each panel corresponds to a different value of $\Delta t_\mathrm{start}^\mathrm{H1}=t_\mathrm{start}^\mathrm{H1}-\bar{t}\,_{\mathrm{peak}}^\mathrm{H1}$, quoted in units of $M$. All $\Delta t_\mathrm{start}^\mathrm{H1}$ values used in panels with dark (light) gold backgrounds are consistent with the median of the $t_{\mathrm{peak}}^\mathrm{H1}$ distribution at $1\sigma$ ($2\sigma$).
In each panel, dashed black, solid red, and solid blue contours correspond to $90\%$ credible level in the \ac{BH} parameters measured using the full \ac{IMR}~\cite{TheLIGOScientific:2016wfe}, $\mathrm{Kerr}_{221}$, and $\mathrm{Kerr}_{220}$ models, respectively.
   }
   \label{fig:Mf_and_af_N1_vs_N0_vary_time}
\end{figure*}

It has long been known that including overtones improves the agreement between ringdown-only fits and the complete gravitational waveforms from perturbed \ac{BH}s. This was first shown by direct integration of the perturbation equations sourced by infalling particles or collapsing matter~\cite{Davis:1971gg,Cunningham:1978zfa,Cunningham:1979px,Ferrari:1981dh} and then, more rigorously, using Green's function techniques~\cite{Leaver:1986gd,Andersson:1996cm,Berti:2006wq,Zhang:2013ksa,Oshita:2021iyn}. Overtones were shown to improve agreement with numerical simulations of collapse~\cite{Stark:1985da}, head-on collisions~\cite{Anninos:1993zj} and quasicircular mergers~\cite{Buonanno:2006ui} leading to \ac{BH} formation, and their omission leads to significant biases in mass and spin estimates~\cite{Berti:2007zu,Baibhav:2017jhs}.
However, standard \ac{QNM} tests often relied only on fundamental modes for two main reasons: overtones are short-lived and difficult to confidently identify in the data~\cite{London:2014cma}, and it is unclear whether multiple overtones have physical meaning or they just happen to phenomenologically fit the nonlinear part of the merger signal~\cite{Buonanno:2006ui,Berti:2007fi}.

Recently, Ref.~\cite{Giesler:2019uxc} showed that including overtones up to $n=7$ in the ringdown model improves the agreement with numerical relativity simulations for all times beyond the time $t_{\mathrm{peak}}$ where $|h_+^2 + h_\times^2|$ has a maximum, claiming that this observation
``implies that the spacetime is well described as a linearly perturbed \ac{BH} with a fixed mass and spin as early as the peak.''
Their study's insistence on an intrinsically linear physical description spurred a sequence of additional investigations, both on the modeling and on the observational side~\cite{Bhagwat:2019dtm,JimenezForteza:2020cve,Bustillo:2020buq,Okounkova:2020vwu,Mourier:2020mwa,Cook:2020otn,Oshita:2021iyn,Dhani:2020nik,Dhani:2021vac,Finch:2021iip,Zertuche:2021xkb}.
If higher overtones can indeed be measured by starting at the peak, the larger ringdown \ac{SNR} would open the door to more precise tests of \ac{GR}. 
This theoretical argument motivated a reanalysis of GW150914. Ref.~\cite{Isi:2019aib} fitted the post-peak waveform with a \ac{QNM} superposition including overtones, and claimed evidence for ``at least one overtone [...] with $3.6 \sigma$ confidence.'' 
The claim seems at odds with Ref.~\cite{Bustillo:2020buq} and with the subsequent LVK analysis~\cite{LIGOScientific:2020tif}, both reporting weak evidence (with a $\log_{10}$-Bayes factor of only $\sim 0.6$) in favor of the ``overtone model'' including both $n=0$ and $n=1$ (henceforth $\mathrm{Kerr}_{221}$) relative to the model including only $n=0$ (henceforth $\mathrm{Kerr}_{220}$).

In this paper we ask whether overtone detection claims in GW150914 data are robust. We use geometrical units $G=c=1$, restoring physical units when needed, and we always quote redshifted \ac{BH} masses as measured in a geocentric reference frame.

\noindent {\bf \em Methods.}
The $\ell=|m|=2$ multipole is largely dominant in GW150914~\cite{Carullo:2019flw,LIGOScientific:2020tif}, so we can ignore higher multipoles and mode-mixing contributions in the general waveform model \eqref{eq:model}. 
The system does not show evidence for antialigned progenitor spins (and more generally, for any non-zero spin), so counterrotating modes can be safely ignored~\cite{LIGOScientific:2020tif, Li:2021wgz}.
We make several assumptions to match as closely as possible the analysis of Ref.~\cite{Isi:2019aib}.  First, we include only one or two \acp{QNM} ($n=0,1$) and assume that all overtones start at the same time $t_{\ell m n}^\mathrm{start} = t_{\mathrm{start}}$.
We fix $(\iota,\phi) = (\pi, 0)\,\mathrm{rad}$, since in our model these parameters are strongly degenerate with the free overtone amplitudes and phases, respectively. Since there is no evidence for misaligned spins in GW150914, we also assume that the waveform amplitudes satisfy $h_{\ell m} = h_{\ell -m}^*$, a good approximation when the progenitor spins are nearly aligned with the orbital angular momentum of the binary.
The strain measured by \ac{GW} detectors is $h_\mathrm{D}(t) = F_+ h_+ + F_\times h_\times$, where the detector pattern functions $F_{+,\times}(\alpha,\delta,\psi)$ depend on the right ascension, declination and polarization angles $\alpha$, $\delta$ and $\psi$~\cite{Maggiore:2007ulw}. 
Following Ref.~\cite{Isi:2019aib} we set $(\alpha,\delta,\psi) = (1.95,-1.27,0.82)\, \mathrm{rad}$.
We fix $t_\mathrm{start}^{\mathrm{H1}}$ in the Hanford detector and compute the starting time in the Livingston detector using a fixed time delay determined from the sky position parameters listed above.
We assume flat priors on all free parameters in the ranges $M_f \in [20,200] \, M_\odot, a_f \in [0, 0.99], A_{2 2 n} \in [0,5\times10^{-20}], \phi_{2 2 n} \in [0,2\pi]$.

\begin{figure*}[thbp]
\centering
\includegraphics[width=\textwidth]{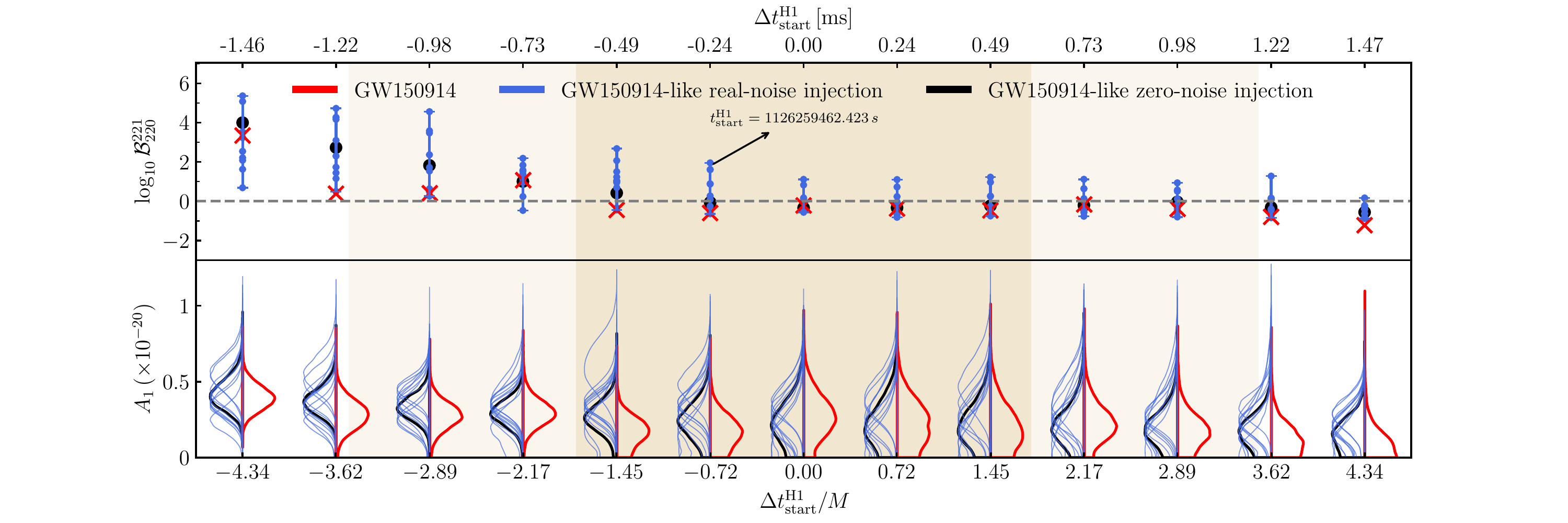}
\caption{
\textit{Top}: Log-Bayes factor $\left(\log_{10}\mathcal{B}_{220}^{221}\right)$ between the $\mathrm{Kerr}_{221}$ and $\mathrm{Kerr}_{220}$ hypotheses as a function of $\Delta t_\mathrm{start}^\mathrm{H1}=t_\mathrm{start}^\mathrm{H1} -\bar{t}\,_{\mathrm{peak}}^\mathrm{H1}$.
For the GW150914 signal (red crosses), $\bar{t}\,_{\mathrm{peak}}^\mathrm{H1}$ is the median of the posterior distribution from the full \ac{IMR} analysis; dark (light) gold bands correspond to the $1\sigma$ ($2\sigma$) uncertainties on the median. For the GW150914-like injections (black), $t_{\mathrm{peak}}^\mathrm{H1}$ is computed from the simulation, and so it is known exactly.
Black dots correspond to a GW150914-like injection in zero noise. The blue dots (and related ``error bars'') are computed by repeating the analysis at each $t_{\mathrm{start}}^{\mathrm{H1}}$ under different realizations of the real detector noise close to the GW150914 trigger.
\textit{Bottom}: Amplitude of the overtone $A_1$ measured for different $t_\mathrm{start}^\mathrm{H1}$. The red (black) curves correspond to the measurement obtained from the GW150914 signal (GW150914-like injection in zero noise). The blue curves are the overtone amplitudes measured on the GW150914-like injection in real noise.}
   \label{fig:BF_and_amp_N1_vs_N0_vary_time}
\end{figure*}

We analyze the ringdown signal using the Bayesian parameter estimation package \pyRing~\cite{pyRing, Carullo:2019flw}, employed by the \ac{LVK} collaboration to perform ringdown-only tests of \ac{GR}.
The \pyRing package relies on the nested sampling algorithm \texttt{cpnest}~\cite{cpnest} (for additional details needed to reproduce our analysis, see the Software section), that allows us to compare alternative hypotheses by computing their relative Bayes factors. 
We use 4096 live points and 4096 maximum Markov Chain (MC) steps, which typically result in $\sim 20000$ independent samples at the end of each of our runs. We have tested the robustness of our results to sampling configurations by repeating the runs close to the peaktime using 10000 live points and MC steps, together with four different random seeds in the instantiations of the nested sampling. All the obtained results are consistently recovered under these changes of settings.
The autocorrelation function (ACF) of the background noise was chosen to be as close as possible to the settings of Ref.~\cite{Isi:2019aib}.
The ACF was computed using a stretch of 64s of data starting at 1126257417s of GPS time (see the Software section for more details). We have verified that ACFs estimated using different data stretches close to the event do not significantly impact our conclusions, in agreement with the hypothesis of wide-sense stationarity of the noise.
The data are appropriately cropped to avoid contamination from earlier stages of the coalescence~\cite{Isi:2021iql}, beginning from the starting time of the analysis and up to a duration of $0.1$\,s.
We analyze publicly available data from GWOSC~\cite{LIGOScientific:2019lzm} with a sampling rate of $16384$\,Hz (the maximum resolution available). This rate, larger than the rate of $2048$\,Hz used in Ref.~\cite{Isi:2019aib}, was chosen to minimize the impact of the time discretization. Repeating the analysis using a rate of $4096$\,Hz left our conclusions unaltered.
When investigating the consequences of slightly changing the analysis settings, we found that the choice of $t_{\mathrm{start}}$ (which has be set equal to $t_{\mathrm{peak}}$ according to the theoretical arguments in~\cite{Giesler:2019uxc}) has by far the largest impact. 
The effect of varying $\psi$, $\iota$ is milder, and it will be discussed in a forthcoming paper~\cite{inprep}, together with the impact of dropping the symmetry assumption on the amplitudes $\hlm$.
Ref.~\cite{Isi:2019aib} assumed $t_{\mathrm{start}} ^\mathrm{H1}= t_{\mathrm{peak}}^\mathrm{H1} = 1126259462.423\, \mathrm{s}$. However the value of $t_{\mathrm{peak}}^\mathrm{H1} $ must be estimated from the data, and as such it is uncertain. 
Fixing it to a specific value can induce systematic biases.
We quantify this uncertainty by reconstructing $t_{\mathrm{peak}}^\mathrm{H1}$ using the posterior distributions of the parameters of GW150914~\cite{O1_samples_release} obtained with the \ac{IMR} waveform model SEOBNRv4~\cite{Bohe:2016gbl} (see the Supplemental Material for details). We check that the reconstruction is robust against waveform systematics by using also the IMRPhenomPv2 waveform model~\cite{Husa:2015iqa, Khan:2015jqa, Hannam:2013oca}.  
In the Hanford detector, the resulting posterior distribution has median $\bar{t}\,_{\mathrm{peak}}^\mathrm{H1} = 1126259462.42323\, \mathrm{s}$ and standard deviation $\sigma = 0.00059\, \mathrm{s}$.
We will vary $t_\mathrm{start}^{\mathrm{H1}}$ within the $\pm 2\sigma$ interval of its posterior distribution.

\noindent {\bf \em Mass and spin estimates.}
In Fig.~\ref{fig:Mf_and_af_N1_vs_N0_vary_time} we show the mass and spin of the GW150914 \ac{BH} remnant estimated using the $\mathrm{Kerr}_{220}$ (blue), $\mathrm{Kerr}_{221}$ (red) and full \ac{IMR} model~\cite{TheLIGOScientific:2016wfe} (dashed black) for 10 selected values of $\Delta t_\mathrm{start}^\mathrm{H1} \equiv t_\mathrm{start}^\mathrm{H1} - \bar{t}\,_{\mathrm{peak}}^\mathrm{H1}$.
For $\Delta t_\mathrm{start}^\mathrm{H1} / M \geq -1.45$, the \ac{IMR} posterior overlaps with both the $\mathrm{Kerr}_{220}$ and $\mathrm{Kerr}_{221}$ models at $90\,\%$ credibility, although the $\mathrm{Kerr}_{221}$ reconstruction peaks closer to the IMR estimate.
The $\mathrm{Kerr}_{221}$ model agrees much better than $\mathrm{Kerr}_{220}$ with the IMR posterior especially when we start fitting before the peak ($\Delta t_\mathrm{start}^\mathrm{H1} / M \leq -2.17$), where such a fit is not well motivated by the overtone model (see Fig.~1 of~\cite{Giesler:2019uxc}). 
The starting time used in Ref.~\cite{Isi:2019aib} corresponds to $\Delta t_{\mathrm{start}}^{\mathrm{H1}}/M= -0.72$ in Fig.~\ref{fig:Mf_and_af_N1_vs_N0_vary_time}. Note that the $\left( M_f, a_f \right)$ measurements obtained with the $\mathrm{Kerr}_{221}$ model overlap with the \ac{GR} prediction even when $\Delta t_\mathrm{start}^\mathrm{H1}/M = -3.62$, outside of the $2\sigma$ confidence interval on the peak location. This is likely due to a combination of two effects: (i) since $\omega_{221}<\omega_{220}$, any overtone model naturally includes a low-frequency component, thus improving the fit to the low-frequency, pre-merger part of the signal; and (ii) the $\mathrm{Kerr}_{221}$ model has a larger number of parameters than the $\mathrm{Kerr}_{220}$ model, thus at low signal-to-noise ratios it can still fit the signal with the values of $\left( M_f, a_f \right)$ determined by the late-time ringdown behavior.

\noindent {\bf \em Bayes factors.}
To quantify the evidence for the presence of an overtone in GW150914, we compare the hypotheses that the data can be described by the $\mathrm{Kerr}_{221}$ vs. $\mathrm{Kerr}_{220}$ models and compute the resulting Bayes factor, $\mathcal{B}_{220}^{221}$.  In the top panel of Fig.~\ref{fig:BF_and_amp_N1_vs_N0_vary_time} we show $\log_{10} \mathcal{B}_{220}^{221}$ (red crosses) for selected valus of $\Delta t_\mathrm{start}^\mathrm{H1}$.  In the bottom panel we show the posterior of the overtone amplitude $A_1 \equiv A_{221}$ for the $\mathrm{Kerr}_{221}$ model (red curves).  When $\Delta t_\mathrm{start}^\mathrm{H1}/M \geq -1.45$, there is no evidence for the overtone in the data ($\log_{10} \mathcal{B}_{220}^{221} < 0$), and the posterior distributions in the bottom panel have significant support for $A_1=0$, hence the $\mathrm{Kerr}_{220}$ model is favored with respect to $\mathrm{Kerr}_{221}$.  We observe significant Bayesian evidence for the presence of the overtone ($\log_{10} \mathcal{B}_{220}^{221} > 2$) only for $\Delta t_\mathrm{start}^\mathrm{H1} / M \leq -4.34$, i.e., well outside of the nominal region of validity of the $\mathrm{Kerr}_{221}$ model.  For $\Delta t_\mathrm{start}^\mathrm{H1}/M= -0.72$, which corresponds to the $t_{\mathrm{peak}}^\mathrm{H1}$ value used in Ref.~\cite{Isi:2019aib}, we find that $\log_{10} \mathcal{B}_{220}^{221} = -0.60$, while the amplitude has large support for zero. At the peak time $A_1$ is maximum away from zero, but there is still some support for zero amplitude. This may lead us to conclude that the overtone is measurable in this ringdown signal. However, both the Bayes factor and $A_1$ {\em decrease} for values of $\Delta t_\mathrm{start}^\mathrm{H1}$ located immediately before and after $\Delta t_\mathrm{start}^\mathrm{H1} / M = 0 $.  Now, the decay time for the overtone in question is $\tau_{221} \approx 1.3 \, \mathrm{ms} \approx 4 M$. If the overtone were measurable, we would expect to find evidence for its presence when changing $t_\mathrm{start}^\mathrm{H1}$ by only $\sim 0.24 \mathrm{ms} \approx 0.72 \, M$.  Since this is not the case, we must consider the hypothesis that the (weak) evidence in favor of an overtone for $\Delta t_\mathrm{start}^\mathrm{H1} / M = 0$ could be driven by a noise fluctuation.

We test this hypothesis by using a synthetic signal (``injection'', in \ac{LVK} jargon) obtained from a numerical solution of the Einstein equations consistent with the GW150914 signal~\cite{Boyle:2019kee} (see the Supplemental Material for details). In this case, $t_{\mathrm{peak}}^\mathrm{H1}$ is known exactly. We analyze the signal using different values of $t_{\mathrm{start}}^{\mathrm{H1}}$, such that $\Delta t_\mathrm{start}^\mathrm{H1}$ is consistent with the values used for the real signal.
For each selected $\Delta t_\mathrm{start}^\mathrm{H1}$, we first perform the analysis described above in the case of the real signal, but we now set the noise realization to zero (``zero-noise'' injection). The resulting parameter distributions will thus have an uncertainty consistent with the actual signal, while eliminating a possible shift of the posterior median due to noise fluctuations coincident with the signal.
The values of $\log_{10} \mathcal{B}_{220}^{221}$ and $A_1$ obtained from this zero-noise injection are shown as black dots and black curves in the upper and lower panels of Fig.~\ref{fig:BF_and_amp_N1_vs_N0_vary_time}. When $\Delta t_\mathrm{start}^\mathrm{H1} / M = 0$ there is no evidence for an overtone ($\log_{10} \mathcal{B}_{220}^{221}=-0.21 < 0$) and $A_1$ has a large support for zero. For the zero-noise injection, the Bayes factor is greater than unity only when $\Delta t_\mathrm{start}^\mathrm{H1}/M\leq  -1.45$, and it generally increases for lower values of $\Delta t_\mathrm{start}^\mathrm{H1}$, similarly to what happens for the real signal. The inferred amplitude of the overtone is consistent with the behavior observed for the Bayes factor, increasing for large negative values of $\Delta t_\mathrm{start}^\mathrm{H1}/M$.

To assess the impact of the detector noise on the measurement of $\log_{10} \mathcal{B}_{220}^{221}$ and $A_1$, for each $\Delta t_\mathrm{start}^\mathrm{H1}$ we repeat the above analysis superposing the simulated signal to $10$ different segments of the real detector noise close to the time of coalescence of GW150914 (see the Supplemental Material).
The resulting Bayes factors are reported as blue dots and related ``error bars'' on $\log_{10} \mathcal{B}_{220}^{221}$: for each time $\Delta t_\mathrm{start}^\mathrm{H1}$, each dot corresponds to a specific noise realisation, while the upper (lower) boundary of the error bar corresponds to the largest (smallest) $\log_{10} \mathcal{B}_{220}^{221}$ obtained from these injections.
The blue curves in the lower panel are the posterior distributions of $A_1$ corresponding to the different noise realisations. These distributions (to be compared with the zero-noise black curves) quantify the impact of noise fluctuations on amplitude measurements.
For $\Delta t_\mathrm{start}^\mathrm{H1} / M = 0$ and neighboring points, the negative values of $\log_{10} \mathcal{B}_{220}^{221}$ measured in the real signal are consistent with the negative values measured in the synthetic signal, if we account for the detector noise. The posterior distributions of $A_1$ shows that a ``favorable'' realization of the detector noise can lead to a measurement of $A_1$ that peaks away from zero (blue curves) -- similarly to the actual signal (red curve) -- although $A_1$ is consistent with zero in the case of the zero-noise injection (black curve). 
We conclude that the mild support for an overtone observed in the amplitude posterior (although never confirmed by the Bayesian evidence) is driven by the detector noise. 

\noindent {\bf \em Discussion.} 
We have performed a Bayesian data analysis of the GW150914 ringdown signal to understand if ringdown overtone detection claims are robust. 
We found no Bayesian evidence in favor of an overtone, nor a significant overtone amplitude measurement in GW150914 data after the waveform peak, where the inclusion of overtones in the ringdown model is expected to improve the agreement with numerical relativity simulations~\cite{Baibhav:2017jhs,Giesler:2019uxc}. 
There is mild support for a nonzero overtone amplitude in the data at the peak, but such support for $A_1 = 0$ is sensitive to changes in the starting time smaller than the overtone damping time. Most importantly, the Bayes factors never favor the detection of an overtone when varying the starting time within the $1 \sigma$ credible region of the peak time reconstruction.
This suggests that the detection is noise-dominated. We verified this hypothesis by performing GW150914-like injections in different segments of the real detector noise. These results differ from Ref.~\cite{Isi:2021iql}, where the impact of the real detector noise and peak time uncertainty were not considered.

For both real and synthetic signals, the evidence for the overtone and the uncertainty on the evidence (as measured by the blue ``error bars'') generally increase for large negative values of $\Delta t_\mathrm{start}^\mathrm{H1}$. The overtone model is not expected to be valid in this region, but the larger number of degrees of freedom in the model can pick up a larger portion of the low-frequency, pre-merger signal power. At the same time, the evidence uncertainty grows dramatically -- spanning up to four orders of magnitude for the earliest times shown in Fig.~\ref{fig:BF_and_amp_N1_vs_N0_vary_time} -- because the poorly constrained model can easily pick up noise fluctuations.

Our results reveal an intrinsic instability of the inference based on such a model. The instability may happen even in the absence of noise, because the mass and spin of the remnant extracted from numerical simulations vary significantly close to the peak of the radiation~\cite{Berti:2007fi,Baibhav:2017jhs,Sberna:2021eui}, and thus the assumption of a linear superposition of \ac{QNM}s starting at the peak can lead to conceptual issues~\cite{Bhagwat:2017tkm,Bhagwat:2019dtm}.  As reported in Table~I of Ref.~\cite{Giesler:2019uxc}, the amplitude of the fundamental mode is stable up to a few parts in $10^3$ under the addition of overtones, but higher overtones have much less stable amplitudes: $A_{221}$ varies by $8\%$, while $A_{223}$ varies by more than $200\%$. This is inconsistent with our understanding of ringdown in the linearized regime, where (by definition) the \ac{QNM} amplitudes should be constant~\cite{Dorband:2006gg,London:2014cma,London:2018gaq,JimenezForteza:2020cve}. This phenomenon was also found in Ref.~\cite{Forteza:2021wfq} over the full nonprecessing parameter space. In the absence of fitting errors for the overtone amplitudes, it is difficult to quantify how much of this variation can be ascribed to the current accuracy of numerical \ac{BH} merger simulations, rather than being due to a time-evolving background. This instability might also explain the incompatibility of the measurement $A_{221} / A_{220} \leq 2$  reported in~\cite{Isi:2019aib, Isi:2021iql}, compared to the predicted value $A_{221} / A_{220} \sim 4$ reported in Table I of~\cite{Giesler:2019uxc}.

A physical parametrization of the overtone amplitudes as a function of the progenitors parameters, similar to the one proposed in Refs.~\cite{London:2014cma,London:2018gaq} for the fundamental modes, may alleviate this problem. However parametrizations of nonspinning binary \ac{BH} mergers find that such a ``global'' fit is not robust under variations of the starting time: see e.g. Figs.~3 and 4 of~\cite{JimenezForteza:2020cve}. 
Overfitting issues are particularly difficult to address. For example, the accuracy of overtone models constructed using \ac{GR} \acp{QNM} can be matched (or even surpassed) by adding ``unphysical'' low-frequency components corresponding to non-GR values of the frequency and damping time~\cite{Bhagwat:2019dtm,Mourier:2020mwa}. Similar ``pseudo-\acp{QNM}'' were introduced in the context of effective-one-body models~\cite{Pan:2011gk,Damour:2014yha,Brito:2018rfr}.

Our results for the Bayes factors are consistent with previous work.
The large number of free parameters in the overtone model introduces an Occam penalty that must be balanced by large \acp{SNR}~\cite{Bustillo:2020buq}.
Even when modeling the overtone amplitudes as functions of the properties of the remnant progenitors, measuring several overtone frequencies may still be impractical: Fisher matrix estimates~\cite{JimenezForteza:2020cve} suggest that it will be easier to obtain evidence for multiple modes using higher angular harmonics rather than overtones. These results are in contrast with the predictions of~\cite{Isi:2021iql}, which employed a different detection criterion. In future work we plan to investigate strategies for a robust modeling and measurement of higher overtones, and to revisit the \ac{BH} spectroscopy horizon estimates of Refs.~\cite{Berti:2016lat,Ota:2021ypb}.

\noindent
{\bf \em Addendum.}
While this paper was under review, some of the authors of~\cite{Isi:2019aib} revisited their original analysis, 
extending it to multiple times around the peak~\cite{Isi:2022mhy}. 
In the Supplemental Material we present a comparison with their publicly available data.
Small differences between the two analyses (i.e., a different sampling algorithm, data sampling rate, and autocorrelation function estimation method) lead to moderately different overtone amplitudes, but we observe broad agreement with our main results. In particular, both sets of posteriors show significant railing against zero within the peak time uncertainty. This comparison does not point to any fundamental discrepancy between the two investigations, and our conclusions are unaltered.

A third independent reanalysis~\cite{Finch:2022ynt} made use of a standard frequency domain approach employed for most of the LVK parameter estimation runs, hence relying on extensively tested algorithms for sampling and estimation of the noise properties.
The authors confirm our main conclusions. They report a ``modest'' ($1.8\sigma$) significance for the detection of an overtone, whereas Ref.~\cite{Isi:2019aib} claimed ``$3.6\sigma$ confidence.'' Perhaps more remarkably, the authors of Ref.~\cite{Finch:2022ynt} find a \textit{negative} Bayes factor in favor of an overtone when marginalizing over all of the relevant uncertainty in the peak strain time. 
Their work confirms that current detection claims depend on subtle data analysis details (such as, e.g., frequency-domain vs. time-domain estimation of the noise properties), which should not have any impact on a robust detection.

\noindent
{\bf \em Acknowledgments.}
We are grateful to Max Isi for help in reproducing the analysis settings of Ref.~\cite{Isi:2019aib}, and to Walter Del Pozzo for valuable comments and suggestions. We thank Vishal Baibhav, Swetha Bhagwat, Juan Calderón Bustillo, Will Farr, Xisco Jiménez-Forteza, Danny Laghi, Lionel London, Paolo Pani, Saul Teukolsky and the Testing GR group of the \ac{LVK} collaboration for stimulating discussions.
R.C. and E.B. are supported by NSF Grants No. PHY-1912550, AST-2006538, PHY-090003 and PHY-20043, and NASA Grants No. 17-ATP17-0225, 19-ATP19-0051 and 20-LPS20-0011. 
This research project was conducted using computational resources at the Maryland Advanced Research Computing Center (MARCC). 
G.C. acknowledges support by the Della Riccia Foundation under an Early Career Scientist Fellowship.
V.C. is a Villum Investigator supported by VILLUM FONDEN (grant no. 37766) and a DNRF Chair supported by the Danish National Research Foundation.
This project has received funding from the European Union's Horizon 2020 research and innovation programme under the Marie Sklodowska-Curie grant agreement No 101007855.
We thank FCT for financial support through Project~No.~UIDB/00099/2020.
We acknowledge financial support provided by FCT/Portugal through grants PTDC/MAT-APL/30043/2017 and PTDC/FIS-AST/7002/2020.
The authors would like to acknowledge networking support by the GWverse COST Action CA16104, ``Black holes, gravitational waves and fundamental physics''.
This material is based upon work supported by NSF's LIGO Laboratory which is a major facility fully funded by the National Science Foundation.\\

\noindent
{\bf \em Software.}
LIGO-Virgo data are interfaced through \texttt{GWpy}~\cite{gwpy}.
Projections onto detectors are computed through \texttt{LALSuite}~\cite{lalsuite}.
The ACFs are computed using the function \texttt{get\_acf} of the \texttt{ringdown} package~\cite{Isi:2021iql}.
The \texttt{pyRing} package is publicly available at: \href{https://git.ligo.org/lscsoft/pyring}{https://git.ligo.org/lscsoft/pyring}.
We use the \texttt{cpnest} version \texttt{0.11.3} and the \pyRing commit \texttt{2b96c569ff663bb71dabe6dae5f4177b79854340} on the master branch.
To allow for reproducibility, we release the configuration file employed for our analysis at the reference time: see~\href{https://github.com/rcotesta/GW150914\_ringdown}{https://github.com/rcotesta/GW150914\_ringdown}.
The other results on observational data can be reproduced by changing the starting time by the amount specified in Fig.~\ref{fig:BF_and_amp_N1_vs_N0_vary_time},
while we give the details needed to reproduce the injections in the Supplemental Material.
This study made use of the open-software \texttt{python} packages: \texttt{corner, cython, h5py, matplotlib, numpy, scipy, seaborn}~\cite{corner, cython, h5py, matplotlib, numpy, scipy, seaborn}.

\bibliography{paper_prl}

\begin{thebibliography}{89}%
\makeatletter
\providecommand \@ifxundefined [1]{%
 \@ifx{#1\undefined}
}%
\providecommand \@ifnum [1]{%
 \ifnum #1\expandafter \@firstoftwo
 \else \expandafter \@secondoftwo
 \fi
}%
\providecommand \@ifx [1]{%
 \ifx #1\expandafter \@firstoftwo
 \else \expandafter \@secondoftwo
 \fi
}%
\providecommand \natexlab [1]{#1}%
\providecommand \enquote  [1]{``#1''}%
\providecommand \bibnamefont  [1]{#1}%
\providecommand \bibfnamefont [1]{#1}%
\providecommand \citenamefont [1]{#1}%
\providecommand \href@noop [0]{\@secondoftwo}%
\providecommand \href [0]{\begingroup \@sanitize@url \@href}%
\providecommand \@href[1]{\@@startlink{#1}\@@href}%
\providecommand \@@href[1]{\endgroup#1\@@endlink}%
\providecommand \@sanitize@url [0]{\catcode `\\12\catcode `\$12\catcode
  `\&12\catcode `\#12\catcode `\^12\catcode `\_12\catcode `\%12\relax}%
\providecommand \@@startlink[1]{}%
\providecommand \@@endlink[0]{}%
\providecommand \url  [0]{\begingroup\@sanitize@url \@url }%
\providecommand \@url [1]{\endgroup\@href {#1}{\urlprefix }}%
\providecommand \urlprefix  [0]{URL }%
\providecommand \Eprint [0]{\href }%
\providecommand \doibase [0]{http://dx.doi.org/}%
\providecommand \selectlanguage [0]{\@gobble}%
\providecommand \bibinfo  [0]{\@secondoftwo}%
\providecommand \bibfield  [0]{\@secondoftwo}%
\providecommand \translation [1]{[#1]}%
\providecommand \BibitemOpen [0]{}%
\providecommand \bibitemStop [0]{}%
\providecommand \bibitemNoStop [0]{.\EOS\space}%
\providecommand \EOS [0]{\spacefactor3000\relax}%
\providecommand \BibitemShut  [1]{\csname bibitem#1\endcsname}%
\let\auto@bib@innerbib\@empty
\bibitem [{\citenamefont {Abbott}\ \emph
  {et~al.}(2016{\natexlab{a}})\citenamefont {Abbott} \emph
  {et~al.}}]{LIGOScientific:2016aoc}%
  \BibitemOpen
  \bibfield  {author} {\bibinfo {author} {\bibfnamefont {B.~P.}\ \bibnamefont
  {Abbott}} \emph {et~al.} (\bibinfo {collaboration} {LIGO Scientific,
  Virgo}),\ }\href {\doibase 10.1103/PhysRevLett.116.061102} {\bibfield
  {journal} {\bibinfo  {journal} {Phys. Rev. Lett.}\ }\textbf {\bibinfo
  {volume} {116}},\ \bibinfo {pages} {061102} (\bibinfo {year}
  {2016}{\natexlab{a}})},\ \Eprint {http://arxiv.org/abs/1602.03837}
  {arXiv:1602.03837 [gr-qc]} \BibitemShut {NoStop}%
\bibitem [{\citenamefont {Aasi}\ \emph {et~al.}(2015)\citenamefont {Aasi} \emph
  {et~al.}}]{LIGOScientific:2014pky}%
  \BibitemOpen
  \bibfield  {author} {\bibinfo {author} {\bibfnamefont {J.}~\bibnamefont
  {Aasi}} \emph {et~al.} (\bibinfo {collaboration} {LIGO Scientific}),\ }\href
  {\doibase 10.1088/0264-9381/32/7/074001} {\bibfield  {journal} {\bibinfo
  {journal} {Class. Quant. Grav.}\ }\textbf {\bibinfo {volume} {32}},\ \bibinfo
  {pages} {074001} (\bibinfo {year} {2015})},\ \Eprint
  {http://arxiv.org/abs/1411.4547} {arXiv:1411.4547 [gr-qc]} \BibitemShut
  {NoStop}%
\bibitem [{\citenamefont {Acernese}\ \emph {et~al.}(2015)\citenamefont
  {Acernese} \emph {et~al.}}]{VIRGO:2014yos}%
  \BibitemOpen
  \bibfield  {author} {\bibinfo {author} {\bibfnamefont {F.}~\bibnamefont
  {Acernese}} \emph {et~al.} (\bibinfo {collaboration} {VIRGO}),\ }\href
  {\doibase 10.1088/0264-9381/32/2/024001} {\bibfield  {journal} {\bibinfo
  {journal} {Class. Quant. Grav.}\ }\textbf {\bibinfo {volume} {32}},\ \bibinfo
  {pages} {024001} (\bibinfo {year} {2015})},\ \Eprint
  {http://arxiv.org/abs/1408.3978} {arXiv:1408.3978 [gr-qc]} \BibitemShut
  {NoStop}%
\bibitem [{\citenamefont {Akutsu}\ \emph {et~al.}(2021)\citenamefont {Akutsu}
  \emph {et~al.}}]{KAGRA:2020tym}%
  \BibitemOpen
  \bibfield  {author} {\bibinfo {author} {\bibfnamefont {T.}~\bibnamefont
  {Akutsu}} \emph {et~al.} (\bibinfo {collaboration} {KAGRA}),\ }\href
  {\doibase 10.1093/ptep/ptaa125} {\bibfield  {journal} {\bibinfo  {journal}
  {PTEP}\ }\textbf {\bibinfo {volume} {2021}},\ \bibinfo {pages} {05A101}
  (\bibinfo {year} {2021})},\ \Eprint {http://arxiv.org/abs/2005.05574}
  {arXiv:2005.05574 [physics.ins-det]} \BibitemShut {NoStop}%
\bibitem [{\citenamefont {Abbott}\ \emph {et~al.}(2019)\citenamefont {Abbott}
  \emph {et~al.}}]{LIGOScientific:2018mvr}%
  \BibitemOpen
  \bibfield  {author} {\bibinfo {author} {\bibfnamefont {B.~P.}\ \bibnamefont
  {Abbott}} \emph {et~al.} (\bibinfo {collaboration} {LIGO Scientific,
  Virgo}),\ }\href {\doibase 10.1103/PhysRevX.9.031040} {\bibfield  {journal}
  {\bibinfo  {journal} {Phys. Rev.}\ }\textbf {\bibinfo {volume} {X9}},\
  \bibinfo {pages} {031040} (\bibinfo {year} {2019})},\ \Eprint
  {http://arxiv.org/abs/1811.12907} {arXiv:1811.12907 [astro-ph.HE]}
  \BibitemShut {NoStop}%
\bibitem [{\citenamefont {Abbott}\ \emph
  {et~al.}(2021{\natexlab{a}})\citenamefont {Abbott} \emph
  {et~al.}}]{LIGOScientific:2020ibl}%
  \BibitemOpen
  \bibfield  {author} {\bibinfo {author} {\bibfnamefont {R.}~\bibnamefont
  {Abbott}} \emph {et~al.} (\bibinfo {collaboration} {LIGO Scientific,
  Virgo}),\ }\href {\doibase 10.1103/PhysRevX.11.021053} {\bibfield  {journal}
  {\bibinfo  {journal} {Phys. Rev. X}\ }\textbf {\bibinfo {volume} {11}},\
  \bibinfo {pages} {021053} (\bibinfo {year} {2021}{\natexlab{a}})},\ \Eprint
  {http://arxiv.org/abs/2010.14527} {arXiv:2010.14527 [gr-qc]} \BibitemShut
  {NoStop}%
\bibitem [{\citenamefont {Abbott}\ \emph
  {et~al.}(2021{\natexlab{b}})\citenamefont {Abbott} \emph
  {et~al.}}]{LIGOScientific:2021usb}%
  \BibitemOpen
  \bibfield  {author} {\bibinfo {author} {\bibfnamefont {R.}~\bibnamefont
  {Abbott}} \emph {et~al.} (\bibinfo {collaboration} {LIGO Scientific,
  VIRGO}),\ }\href@noop {} {\  (\bibinfo {year} {2021}{\natexlab{b}})},\
  \Eprint {http://arxiv.org/abs/2108.01045} {arXiv:2108.01045 [gr-qc]}
  \BibitemShut {NoStop}%
\bibitem [{\citenamefont {Abbott}\ \emph
  {et~al.}(2021{\natexlab{c}})\citenamefont {Abbott} \emph
  {et~al.}}]{LIGOScientific:2021djp}%
  \BibitemOpen
  \bibfield  {author} {\bibinfo {author} {\bibfnamefont {R.}~\bibnamefont
  {Abbott}} \emph {et~al.} (\bibinfo {collaboration} {LIGO Scientific, VIRGO,
  KAGRA}),\ }\href@noop {} {\  (\bibinfo {year} {2021}{\natexlab{c}})},\
  \Eprint {http://arxiv.org/abs/2111.03606} {arXiv:2111.03606 [gr-qc]}
  \BibitemShut {NoStop}%
\bibitem [{\citenamefont {Nitz}\ \emph {et~al.}(2019)\citenamefont {Nitz},
  \citenamefont {Capano}, \citenamefont {Nielsen}, \citenamefont {Reyes},
  \citenamefont {White}, \citenamefont {Brown},\ and\ \citenamefont
  {Krishnan}}]{Nitz:2018imz}%
  \BibitemOpen
  \bibfield  {author} {\bibinfo {author} {\bibfnamefont {A.~H.}\ \bibnamefont
  {Nitz}}, \bibinfo {author} {\bibfnamefont {C.}~\bibnamefont {Capano}},
  \bibinfo {author} {\bibfnamefont {A.~B.}\ \bibnamefont {Nielsen}}, \bibinfo
  {author} {\bibfnamefont {S.}~\bibnamefont {Reyes}}, \bibinfo {author}
  {\bibfnamefont {R.}~\bibnamefont {White}}, \bibinfo {author} {\bibfnamefont
  {D.~A.}\ \bibnamefont {Brown}}, \ and\ \bibinfo {author} {\bibfnamefont
  {B.}~\bibnamefont {Krishnan}},\ }\href {\doibase 10.3847/1538-4357/ab0108}
  {\bibfield  {journal} {\bibinfo  {journal} {Astrophys. J.}\ }\textbf
  {\bibinfo {volume} {872}},\ \bibinfo {pages} {195} (\bibinfo {year}
  {2019})},\ \Eprint {http://arxiv.org/abs/1811.01921} {arXiv:1811.01921
  [gr-qc]} \BibitemShut {NoStop}%
\bibitem [{\citenamefont {Nitz}\ \emph {et~al.}(2020)\citenamefont {Nitz},
  \citenamefont {Dent}, \citenamefont {Davies}, \citenamefont {Kumar},
  \citenamefont {Capano}, \citenamefont {Harry}, \citenamefont {Mozzon},
  \citenamefont {Nuttall}, \citenamefont {Lundgren},\ and\ \citenamefont
  {T\'apai}}]{Nitz:2020oeq}%
  \BibitemOpen
  \bibfield  {author} {\bibinfo {author} {\bibfnamefont {A.~H.}\ \bibnamefont
  {Nitz}}, \bibinfo {author} {\bibfnamefont {T.}~\bibnamefont {Dent}}, \bibinfo
  {author} {\bibfnamefont {G.~S.}\ \bibnamefont {Davies}}, \bibinfo {author}
  {\bibfnamefont {S.}~\bibnamefont {Kumar}}, \bibinfo {author} {\bibfnamefont
  {C.~D.}\ \bibnamefont {Capano}}, \bibinfo {author} {\bibfnamefont
  {I.}~\bibnamefont {Harry}}, \bibinfo {author} {\bibfnamefont
  {S.}~\bibnamefont {Mozzon}}, \bibinfo {author} {\bibfnamefont
  {L.}~\bibnamefont {Nuttall}}, \bibinfo {author} {\bibfnamefont
  {A.}~\bibnamefont {Lundgren}}, \ and\ \bibinfo {author} {\bibfnamefont
  {M.}~\bibnamefont {T\'apai}},\ }\href {\doibase 10.3847/1538-4357/ab733f}
  {\bibfield  {journal} {\bibinfo  {journal} {Astrophys. J.}\ }\textbf
  {\bibinfo {volume} {891}},\ \bibinfo {pages} {123} (\bibinfo {year}
  {2020})},\ \Eprint {http://arxiv.org/abs/1910.05331} {arXiv:1910.05331
  [astro-ph.HE]} \BibitemShut {NoStop}%
\bibitem [{\citenamefont {Nitz}\ \emph {et~al.}(2021)\citenamefont {Nitz},
  \citenamefont {Capano}, \citenamefont {Kumar}, \citenamefont {Wang},
  \citenamefont {Kastha}, \citenamefont {Sch\"afer}, \citenamefont
  {Dhurkunde},\ and\ \citenamefont {Cabero}}]{Nitz:2021uxj}%
  \BibitemOpen
  \bibfield  {author} {\bibinfo {author} {\bibfnamefont {A.~H.}\ \bibnamefont
  {Nitz}}, \bibinfo {author} {\bibfnamefont {C.~D.}\ \bibnamefont {Capano}},
  \bibinfo {author} {\bibfnamefont {S.}~\bibnamefont {Kumar}}, \bibinfo
  {author} {\bibfnamefont {Y.-F.}\ \bibnamefont {Wang}}, \bibinfo {author}
  {\bibfnamefont {S.}~\bibnamefont {Kastha}}, \bibinfo {author} {\bibfnamefont
  {M.}~\bibnamefont {Sch\"afer}}, \bibinfo {author} {\bibfnamefont
  {R.}~\bibnamefont {Dhurkunde}}, \ and\ \bibinfo {author} {\bibfnamefont
  {M.}~\bibnamefont {Cabero}},\ }\href {\doibase 10.3847/1538-4357/ac1c03}
  {\bibfield  {journal} {\bibinfo  {journal} {Astrophys. J.}\ }\textbf
  {\bibinfo {volume} {922}},\ \bibinfo {pages} {76} (\bibinfo {year} {2021})},\
  \Eprint {http://arxiv.org/abs/2105.09151} {arXiv:2105.09151 [astro-ph.HE]}
  \BibitemShut {NoStop}%
\bibitem [{\citenamefont {Venumadhav}\ \emph {et~al.}(2020)\citenamefont
  {Venumadhav}, \citenamefont {Zackay}, \citenamefont {Roulet}, \citenamefont
  {Dai},\ and\ \citenamefont {Zaldarriaga}}]{Venumadhav:2019lyq}%
  \BibitemOpen
  \bibfield  {author} {\bibinfo {author} {\bibfnamefont {T.}~\bibnamefont
  {Venumadhav}}, \bibinfo {author} {\bibfnamefont {B.}~\bibnamefont {Zackay}},
  \bibinfo {author} {\bibfnamefont {J.}~\bibnamefont {Roulet}}, \bibinfo
  {author} {\bibfnamefont {L.}~\bibnamefont {Dai}}, \ and\ \bibinfo {author}
  {\bibfnamefont {M.}~\bibnamefont {Zaldarriaga}},\ }\href {\doibase
  10.1103/PhysRevD.101.083030} {\bibfield  {journal} {\bibinfo  {journal}
  {Phys. Rev. D}\ }\textbf {\bibinfo {volume} {101}},\ \bibinfo {pages}
  {083030} (\bibinfo {year} {2020})},\ \Eprint
  {http://arxiv.org/abs/1904.07214} {arXiv:1904.07214 [astro-ph.HE]}
  \BibitemShut {NoStop}%
\bibitem [{\citenamefont {Zackay}\ \emph {et~al.}(2021)\citenamefont {Zackay},
  \citenamefont {Dai}, \citenamefont {Venumadhav}, \citenamefont {Roulet},\
  and\ \citenamefont {Zaldarriaga}}]{Zackay:2019btq}%
  \BibitemOpen
  \bibfield  {author} {\bibinfo {author} {\bibfnamefont {B.}~\bibnamefont
  {Zackay}}, \bibinfo {author} {\bibfnamefont {L.}~\bibnamefont {Dai}},
  \bibinfo {author} {\bibfnamefont {T.}~\bibnamefont {Venumadhav}}, \bibinfo
  {author} {\bibfnamefont {J.}~\bibnamefont {Roulet}}, \ and\ \bibinfo {author}
  {\bibfnamefont {M.}~\bibnamefont {Zaldarriaga}},\ }\href {\doibase
  10.1103/PhysRevD.104.063030} {\bibfield  {journal} {\bibinfo  {journal}
  {Phys. Rev. D}\ }\textbf {\bibinfo {volume} {104}},\ \bibinfo {pages}
  {063030} (\bibinfo {year} {2021})},\ \Eprint
  {http://arxiv.org/abs/1910.09528} {arXiv:1910.09528 [astro-ph.HE]}
  \BibitemShut {NoStop}%
\bibitem [{\citenamefont {Abbott}\ \emph
  {et~al.}(2021{\natexlab{d}})\citenamefont {Abbott} \emph
  {et~al.}}]{LIGOScientific:2021aug}%
  \BibitemOpen
  \bibfield  {author} {\bibinfo {author} {\bibfnamefont {R.}~\bibnamefont
  {Abbott}} \emph {et~al.} (\bibinfo {collaboration} {LIGO Scientific, VIRGO,
  KAGRA}),\ }\href@noop {} {\  (\bibinfo {year} {2021}{\natexlab{d}})},\
  \Eprint {http://arxiv.org/abs/2111.03604} {arXiv:2111.03604 [astro-ph.CO]}
  \BibitemShut {NoStop}%
\bibitem [{\citenamefont {Abbott}\ \emph
  {et~al.}(2021{\natexlab{e}})\citenamefont {Abbott} \emph
  {et~al.}}]{LIGOScientific:2021psn}%
  \BibitemOpen
  \bibfield  {author} {\bibinfo {author} {\bibfnamefont {R.}~\bibnamefont
  {Abbott}} \emph {et~al.} (\bibinfo {collaboration} {LIGO Scientific, VIRGO,
  KAGRA}),\ }\href@noop {} {\  (\bibinfo {year} {2021}{\natexlab{e}})},\
  \Eprint {http://arxiv.org/abs/2111.03634} {arXiv:2111.03634 [astro-ph.HE]}
  \BibitemShut {NoStop}%
\bibitem [{\citenamefont {Chatziioannou}(2020)}]{Chatziioannou:2020pqz}%
  \BibitemOpen
  \bibfield  {author} {\bibinfo {author} {\bibfnamefont {K.}~\bibnamefont
  {Chatziioannou}},\ }\href {\doibase 10.1007/s10714-020-02754-3} {\bibfield
  {journal} {\bibinfo  {journal} {Gen. Rel. Grav.}\ }\textbf {\bibinfo {volume}
  {52}},\ \bibinfo {pages} {109} (\bibinfo {year} {2020})},\ \Eprint
  {http://arxiv.org/abs/2006.03168} {arXiv:2006.03168 [gr-qc]} \BibitemShut
  {NoStop}%
\bibitem [{\citenamefont {Abbott}\ \emph
  {et~al.}(2021{\natexlab{f}})\citenamefont {Abbott} \emph
  {et~al.}}]{LIGOScientific:2020tif}%
  \BibitemOpen
  \bibfield  {author} {\bibinfo {author} {\bibfnamefont {R.}~\bibnamefont
  {Abbott}} \emph {et~al.} (\bibinfo {collaboration} {LIGO Scientific,
  Virgo}),\ }\href {\doibase 10.1103/PhysRevD.103.122002} {\bibfield  {journal}
  {\bibinfo  {journal} {Phys. Rev. D}\ }\textbf {\bibinfo {volume} {103}},\
  \bibinfo {pages} {122002} (\bibinfo {year} {2021}{\natexlab{f}})},\ \Eprint
  {http://arxiv.org/abs/2010.14529} {arXiv:2010.14529 [gr-qc]} \BibitemShut
  {NoStop}%
\bibitem [{\citenamefont {Press}(1971)}]{Press:1971wr}%
  \BibitemOpen
  \bibfield  {author} {\bibinfo {author} {\bibfnamefont {W.~H.}\ \bibnamefont
  {Press}},\ }\href {\doibase 10.1086/180849} {\bibfield  {journal} {\bibinfo
  {journal} {Astrophys. J. Lett.}\ }\textbf {\bibinfo {volume} {170}},\
  \bibinfo {pages} {L105} (\bibinfo {year} {1971})}\BibitemShut {NoStop}%
\bibitem [{\citenamefont {Chandrasekhar}\ and\ \citenamefont
  {Detweiler}(1975)}]{Chandrasekhar:1975zza}%
  \BibitemOpen
  \bibfield  {author} {\bibinfo {author} {\bibfnamefont {S.}~\bibnamefont
  {Chandrasekhar}}\ and\ \bibinfo {author} {\bibfnamefont {S.~L.}\ \bibnamefont
  {Detweiler}},\ }\href {\doibase 10.1098/rspa.1975.0112} {\bibfield  {journal}
  {\bibinfo  {journal} {Proc. Roy. Soc. Lond. A}\ }\textbf {\bibinfo {volume}
  {344}},\ \bibinfo {pages} {441} (\bibinfo {year} {1975})}\BibitemShut
  {NoStop}%
\bibitem [{\citenamefont {Detweiler}(1980)}]{Detweiler:1980gk}%
  \BibitemOpen
  \bibfield  {author} {\bibinfo {author} {\bibfnamefont {S.~L.}\ \bibnamefont
  {Detweiler}},\ }\href {\doibase 10.1086/158109} {\bibfield  {journal}
  {\bibinfo  {journal} {Astrophys. J.}\ }\textbf {\bibinfo {volume} {239}},\
  \bibinfo {pages} {292} (\bibinfo {year} {1980})}\BibitemShut {NoStop}%
\bibitem [{\citenamefont {Kokkotas}\ and\ \citenamefont
  {Schmidt}(1999)}]{Kokkotas:1999bd}%
  \BibitemOpen
  \bibfield  {author} {\bibinfo {author} {\bibfnamefont {K.~D.}\ \bibnamefont
  {Kokkotas}}\ and\ \bibinfo {author} {\bibfnamefont {B.~G.}\ \bibnamefont
  {Schmidt}},\ }\href {\doibase 10.12942/lrr-1999-2} {\bibfield  {journal}
  {\bibinfo  {journal} {Living Rev. Rel.}\ }\textbf {\bibinfo {volume} {2}},\
  \bibinfo {pages} {2} (\bibinfo {year} {1999})},\ \Eprint
  {http://arxiv.org/abs/gr-qc/9909058} {arXiv:gr-qc/9909058} \BibitemShut
  {NoStop}%
\bibitem [{\citenamefont {Dreyer}\ \emph {et~al.}(2004)\citenamefont {Dreyer},
  \citenamefont {Kelly}, \citenamefont {Krishnan}, \citenamefont {Finn},
  \citenamefont {Garrison},\ and\ \citenamefont
  {Lopez-Aleman}}]{Dreyer:2003bv}%
  \BibitemOpen
  \bibfield  {author} {\bibinfo {author} {\bibfnamefont {O.}~\bibnamefont
  {Dreyer}}, \bibinfo {author} {\bibfnamefont {B.~J.}\ \bibnamefont {Kelly}},
  \bibinfo {author} {\bibfnamefont {B.}~\bibnamefont {Krishnan}}, \bibinfo
  {author} {\bibfnamefont {L.~S.}\ \bibnamefont {Finn}}, \bibinfo {author}
  {\bibfnamefont {D.}~\bibnamefont {Garrison}}, \ and\ \bibinfo {author}
  {\bibfnamefont {R.}~\bibnamefont {Lopez-Aleman}},\ }\href {\doibase
  10.1088/0264-9381/21/4/003} {\bibfield  {journal} {\bibinfo  {journal}
  {Class. Quant. Grav.}\ }\textbf {\bibinfo {volume} {21}},\ \bibinfo {pages}
  {787} (\bibinfo {year} {2004})},\ \Eprint
  {http://arxiv.org/abs/gr-qc/0309007} {arXiv:gr-qc/0309007} \BibitemShut
  {NoStop}%
\bibitem [{\citenamefont {Berti}\ \emph {et~al.}(2006)\citenamefont {Berti},
  \citenamefont {Cardoso},\ and\ \citenamefont {Will}}]{Berti:2005ys}%
  \BibitemOpen
  \bibfield  {author} {\bibinfo {author} {\bibfnamefont {E.}~\bibnamefont
  {Berti}}, \bibinfo {author} {\bibfnamefont {V.}~\bibnamefont {Cardoso}}, \
  and\ \bibinfo {author} {\bibfnamefont {C.~M.}\ \bibnamefont {Will}},\ }\href
  {\doibase 10.1103/PhysRevD.73.064030} {\bibfield  {journal} {\bibinfo
  {journal} {Phys. Rev. D}\ }\textbf {\bibinfo {volume} {73}},\ \bibinfo
  {pages} {064030} (\bibinfo {year} {2006})},\ \Eprint
  {http://arxiv.org/abs/gr-qc/0512160} {arXiv:gr-qc/0512160} \BibitemShut
  {NoStop}%
\bibitem [{\citenamefont {Berti}\ \emph {et~al.}(2009)\citenamefont {Berti},
  \citenamefont {Cardoso},\ and\ \citenamefont {Starinets}}]{Berti:2009kk}%
  \BibitemOpen
  \bibfield  {author} {\bibinfo {author} {\bibfnamefont {E.}~\bibnamefont
  {Berti}}, \bibinfo {author} {\bibfnamefont {V.}~\bibnamefont {Cardoso}}, \
  and\ \bibinfo {author} {\bibfnamefont {A.~O.}\ \bibnamefont {Starinets}},\
  }\href {\doibase 10.1088/0264-9381/26/16/163001} {\bibfield  {journal}
  {\bibinfo  {journal} {Class. Quant. Grav.}\ }\textbf {\bibinfo {volume}
  {26}},\ \bibinfo {pages} {163001} (\bibinfo {year} {2009})},\ \Eprint
  {http://arxiv.org/abs/0905.2975} {arXiv:0905.2975 [gr-qc]} \BibitemShut
  {NoStop}%
\bibitem [{\citenamefont {Flanagan}\ and\ \citenamefont
  {Hughes}(1998)}]{Flanagan:1997sx}%
  \BibitemOpen
  \bibfield  {author} {\bibinfo {author} {\bibfnamefont {E.~E.}\ \bibnamefont
  {Flanagan}}\ and\ \bibinfo {author} {\bibfnamefont {S.~A.}\ \bibnamefont
  {Hughes}},\ }\href {\doibase 10.1103/PhysRevD.57.4535} {\bibfield  {journal}
  {\bibinfo  {journal} {Phys. Rev. D}\ }\textbf {\bibinfo {volume} {57}},\
  \bibinfo {pages} {4535} (\bibinfo {year} {1998})},\ \Eprint
  {http://arxiv.org/abs/gr-qc/9701039} {arXiv:gr-qc/9701039} \BibitemShut
  {NoStop}%
\bibitem [{\citenamefont {Buonanno}\ \emph {et~al.}(2007)\citenamefont
  {Buonanno}, \citenamefont {Cook},\ and\ \citenamefont
  {Pretorius}}]{Buonanno:2006ui}%
  \BibitemOpen
  \bibfield  {author} {\bibinfo {author} {\bibfnamefont {A.}~\bibnamefont
  {Buonanno}}, \bibinfo {author} {\bibfnamefont {G.~B.}\ \bibnamefont {Cook}},
  \ and\ \bibinfo {author} {\bibfnamefont {F.}~\bibnamefont {Pretorius}},\
  }\href {\doibase 10.1103/PhysRevD.75.124018} {\bibfield  {journal} {\bibinfo
  {journal} {Phys. Rev. D}\ }\textbf {\bibinfo {volume} {75}},\ \bibinfo
  {pages} {124018} (\bibinfo {year} {2007})},\ \Eprint
  {http://arxiv.org/abs/gr-qc/0610122} {arXiv:gr-qc/0610122} \BibitemShut
  {NoStop}%
\bibitem [{\citenamefont {Berti}\ \emph
  {et~al.}(2007{\natexlab{a}})\citenamefont {Berti}, \citenamefont {Cardoso},
  \citenamefont {Gonzalez}, \citenamefont {Sperhake}, \citenamefont {Hannam},
  \citenamefont {Husa},\ and\ \citenamefont {Bruegmann}}]{Berti:2007fi}%
  \BibitemOpen
  \bibfield  {author} {\bibinfo {author} {\bibfnamefont {E.}~\bibnamefont
  {Berti}}, \bibinfo {author} {\bibfnamefont {V.}~\bibnamefont {Cardoso}},
  \bibinfo {author} {\bibfnamefont {J.~A.}\ \bibnamefont {Gonzalez}}, \bibinfo
  {author} {\bibfnamefont {U.}~\bibnamefont {Sperhake}}, \bibinfo {author}
  {\bibfnamefont {M.}~\bibnamefont {Hannam}}, \bibinfo {author} {\bibfnamefont
  {S.}~\bibnamefont {Husa}}, \ and\ \bibinfo {author} {\bibfnamefont
  {B.}~\bibnamefont {Bruegmann}},\ }\href {\doibase 10.1103/PhysRevD.76.064034}
  {\bibfield  {journal} {\bibinfo  {journal} {Phys. Rev. D}\ }\textbf {\bibinfo
  {volume} {76}},\ \bibinfo {pages} {064034} (\bibinfo {year}
  {2007}{\natexlab{a}})},\ \Eprint {http://arxiv.org/abs/gr-qc/0703053}
  {arXiv:gr-qc/0703053} \BibitemShut {NoStop}%
\bibitem [{\citenamefont {Abbott}\ \emph
  {et~al.}(2016{\natexlab{b}})\citenamefont {Abbott} \emph
  {et~al.}}]{TheLIGOScientific:2016wfe}%
  \BibitemOpen
  \bibfield  {author} {\bibinfo {author} {\bibfnamefont {B.~P.}\ \bibnamefont
  {Abbott}} \emph {et~al.} (\bibinfo {collaboration} {LIGO Scientific,
  Virgo}),\ }\href {\doibase 10.1103/PhysRevLett.116.241102} {\bibfield
  {journal} {\bibinfo  {journal} {Phys. Rev. Lett.}\ }\textbf {\bibinfo
  {volume} {116}},\ \bibinfo {pages} {241102} (\bibinfo {year}
  {2016}{\natexlab{b}})},\ \Eprint {http://arxiv.org/abs/1602.03840}
  {arXiv:1602.03840 [gr-qc]} \BibitemShut {NoStop}%
\bibitem [{\citenamefont {Davis}\ \emph {et~al.}(1971)\citenamefont {Davis},
  \citenamefont {Ruffini}, \citenamefont {Press},\ and\ \citenamefont
  {Price}}]{Davis:1971gg}%
  \BibitemOpen
  \bibfield  {author} {\bibinfo {author} {\bibfnamefont {M.}~\bibnamefont
  {Davis}}, \bibinfo {author} {\bibfnamefont {R.}~\bibnamefont {Ruffini}},
  \bibinfo {author} {\bibfnamefont {W.~H.}\ \bibnamefont {Press}}, \ and\
  \bibinfo {author} {\bibfnamefont {R.~H.}\ \bibnamefont {Price}},\ }\href
  {\doibase 10.1103/PhysRevLett.27.1466} {\bibfield  {journal} {\bibinfo
  {journal} {Phys. Rev. Lett.}\ }\textbf {\bibinfo {volume} {27}},\ \bibinfo
  {pages} {1466} (\bibinfo {year} {1971})}\BibitemShut {NoStop}%
\bibitem [{\citenamefont {Cunningham}\ \emph {et~al.}(1978)\citenamefont
  {Cunningham}, \citenamefont {Price},\ and\ \citenamefont
  {Moncrief}}]{Cunningham:1978zfa}%
  \BibitemOpen
  \bibfield  {author} {\bibinfo {author} {\bibfnamefont {C.~T.}\ \bibnamefont
  {Cunningham}}, \bibinfo {author} {\bibfnamefont {R.~H.}\ \bibnamefont
  {Price}}, \ and\ \bibinfo {author} {\bibfnamefont {V.}~\bibnamefont
  {Moncrief}},\ }\href {\doibase 10.1086/156413} {\bibfield  {journal}
  {\bibinfo  {journal} {Astrophys. J.}\ }\textbf {\bibinfo {volume} {224}},\
  \bibinfo {pages} {643} (\bibinfo {year} {1978})}\BibitemShut {NoStop}%
\bibitem [{\citenamefont {Cunningham}\ \emph {et~al.}(1979)\citenamefont
  {Cunningham}, \citenamefont {Price},\ and\ \citenamefont
  {Moncrief}}]{Cunningham:1979px}%
  \BibitemOpen
  \bibfield  {author} {\bibinfo {author} {\bibfnamefont {C.~T.}\ \bibnamefont
  {Cunningham}}, \bibinfo {author} {\bibfnamefont {R.~H.}\ \bibnamefont
  {Price}}, \ and\ \bibinfo {author} {\bibfnamefont {V.}~\bibnamefont
  {Moncrief}},\ }\href {\doibase 10.1086/157147} {\bibfield  {journal}
  {\bibinfo  {journal} {Astrophys. J.}\ }\textbf {\bibinfo {volume} {230}},\
  \bibinfo {pages} {870} (\bibinfo {year} {1979})}\BibitemShut {NoStop}%
\bibitem [{\citenamefont {Ferrari}\ and\ \citenamefont
  {Ruffini}(1981)}]{Ferrari:1981dh}%
  \BibitemOpen
  \bibfield  {author} {\bibinfo {author} {\bibfnamefont {V.}~\bibnamefont
  {Ferrari}}\ and\ \bibinfo {author} {\bibfnamefont {R.}~\bibnamefont
  {Ruffini}},\ }\href {\doibase 10.1016/0370-2693(81)90930-8} {\bibfield
  {journal} {\bibinfo  {journal} {Phys. Lett. B}\ }\textbf {\bibinfo {volume}
  {98}},\ \bibinfo {pages} {381} (\bibinfo {year} {1981})}\BibitemShut
  {NoStop}%
\bibitem [{\citenamefont {Leaver}(1986)}]{Leaver:1986gd}%
  \BibitemOpen
  \bibfield  {author} {\bibinfo {author} {\bibfnamefont {E.~W.}\ \bibnamefont
  {Leaver}},\ }\href {\doibase 10.1103/PhysRevD.34.384} {\bibfield  {journal}
  {\bibinfo  {journal} {Phys. Rev. D}\ }\textbf {\bibinfo {volume} {34}},\
  \bibinfo {pages} {384} (\bibinfo {year} {1986})}\BibitemShut {NoStop}%
\bibitem [{\citenamefont {Andersson}(1997)}]{Andersson:1996cm}%
  \BibitemOpen
  \bibfield  {author} {\bibinfo {author} {\bibfnamefont {N.}~\bibnamefont
  {Andersson}},\ }\href {\doibase 10.1103/PhysRevD.55.468} {\bibfield
  {journal} {\bibinfo  {journal} {Phys. Rev. D}\ }\textbf {\bibinfo {volume}
  {55}},\ \bibinfo {pages} {468} (\bibinfo {year} {1997})},\ \Eprint
  {http://arxiv.org/abs/gr-qc/9607064} {arXiv:gr-qc/9607064} \BibitemShut
  {NoStop}%
\bibitem [{\citenamefont {Berti}\ and\ \citenamefont
  {Cardoso}(2006)}]{Berti:2006wq}%
  \BibitemOpen
  \bibfield  {author} {\bibinfo {author} {\bibfnamefont {E.}~\bibnamefont
  {Berti}}\ and\ \bibinfo {author} {\bibfnamefont {V.}~\bibnamefont
  {Cardoso}},\ }\href {\doibase 10.1103/PhysRevD.74.104020} {\bibfield
  {journal} {\bibinfo  {journal} {Phys. Rev. D}\ }\textbf {\bibinfo {volume}
  {74}},\ \bibinfo {pages} {104020} (\bibinfo {year} {2006})},\ \Eprint
  {http://arxiv.org/abs/gr-qc/0605118} {arXiv:gr-qc/0605118} \BibitemShut
  {NoStop}%
\bibitem [{\citenamefont {Zhang}\ \emph {et~al.}(2013)\citenamefont {Zhang},
  \citenamefont {Berti},\ and\ \citenamefont {Cardoso}}]{Zhang:2013ksa}%
  \BibitemOpen
  \bibfield  {author} {\bibinfo {author} {\bibfnamefont {Z.}~\bibnamefont
  {Zhang}}, \bibinfo {author} {\bibfnamefont {E.}~\bibnamefont {Berti}}, \ and\
  \bibinfo {author} {\bibfnamefont {V.}~\bibnamefont {Cardoso}},\ }\href
  {\doibase 10.1103/PhysRevD.88.044018} {\bibfield  {journal} {\bibinfo
  {journal} {Phys. Rev. D}\ }\textbf {\bibinfo {volume} {88}},\ \bibinfo
  {pages} {044018} (\bibinfo {year} {2013})},\ \Eprint
  {http://arxiv.org/abs/1305.4306} {arXiv:1305.4306 [gr-qc]} \BibitemShut
  {NoStop}%
\bibitem [{\citenamefont {Oshita}(2021)}]{Oshita:2021iyn}%
  \BibitemOpen
  \bibfield  {author} {\bibinfo {author} {\bibfnamefont {N.}~\bibnamefont
  {Oshita}},\ }\href@noop {} {\  (\bibinfo {year} {2021})},\ \Eprint
  {http://arxiv.org/abs/2109.09757} {arXiv:2109.09757 [gr-qc]} \BibitemShut
  {NoStop}%
\bibitem [{\citenamefont {Stark}\ and\ \citenamefont
  {Piran}(1985)}]{Stark:1985da}%
  \BibitemOpen
  \bibfield  {author} {\bibinfo {author} {\bibfnamefont {R.~F.}\ \bibnamefont
  {Stark}}\ and\ \bibinfo {author} {\bibfnamefont {T.}~\bibnamefont {Piran}},\
  }\href {\doibase 10.1103/PhysRevLett.55.891} {\bibfield  {journal} {\bibinfo
  {journal} {Phys. Rev. Lett.}\ }\textbf {\bibinfo {volume} {55}},\ \bibinfo
  {pages} {891} (\bibinfo {year} {1985})},\ \bibinfo {note} {[Erratum:
  Phys.Rev.Lett. 56, 97 (1986)]}\BibitemShut {NoStop}%
\bibitem [{\citenamefont {Anninos}\ \emph {et~al.}(1993)\citenamefont
  {Anninos}, \citenamefont {Hobill}, \citenamefont {Seidel}, \citenamefont
  {Smarr},\ and\ \citenamefont {Suen}}]{Anninos:1993zj}%
  \BibitemOpen
  \bibfield  {author} {\bibinfo {author} {\bibfnamefont {P.}~\bibnamefont
  {Anninos}}, \bibinfo {author} {\bibfnamefont {D.}~\bibnamefont {Hobill}},
  \bibinfo {author} {\bibfnamefont {E.}~\bibnamefont {Seidel}}, \bibinfo
  {author} {\bibfnamefont {L.}~\bibnamefont {Smarr}}, \ and\ \bibinfo {author}
  {\bibfnamefont {W.-M.}\ \bibnamefont {Suen}},\ }\href {\doibase
  10.1103/PhysRevLett.71.2851} {\bibfield  {journal} {\bibinfo  {journal}
  {Phys. Rev. Lett.}\ }\textbf {\bibinfo {volume} {71}},\ \bibinfo {pages}
  {2851} (\bibinfo {year} {1993})},\ \Eprint
  {http://arxiv.org/abs/gr-qc/9309016} {arXiv:gr-qc/9309016} \BibitemShut
  {NoStop}%
\bibitem [{\citenamefont {Berti}\ \emph
  {et~al.}(2007{\natexlab{b}})\citenamefont {Berti}, \citenamefont {Cardoso},
  \citenamefont {Cardoso},\ and\ \citenamefont {Cavaglia}}]{Berti:2007zu}%
  \BibitemOpen
  \bibfield  {author} {\bibinfo {author} {\bibfnamefont {E.}~\bibnamefont
  {Berti}}, \bibinfo {author} {\bibfnamefont {J.}~\bibnamefont {Cardoso}},
  \bibinfo {author} {\bibfnamefont {V.}~\bibnamefont {Cardoso}}, \ and\
  \bibinfo {author} {\bibfnamefont {M.}~\bibnamefont {Cavaglia}},\ }\href
  {\doibase 10.1103/PhysRevD.76.104044} {\bibfield  {journal} {\bibinfo
  {journal} {Phys. Rev. D}\ }\textbf {\bibinfo {volume} {76}},\ \bibinfo
  {pages} {104044} (\bibinfo {year} {2007}{\natexlab{b}})},\ \Eprint
  {http://arxiv.org/abs/0707.1202} {arXiv:0707.1202 [gr-qc]} \BibitemShut
  {NoStop}%
\bibitem [{\citenamefont {Baibhav}\ \emph {et~al.}(2018)\citenamefont
  {Baibhav}, \citenamefont {Berti}, \citenamefont {Cardoso},\ and\
  \citenamefont {Khanna}}]{Baibhav:2017jhs}%
  \BibitemOpen
  \bibfield  {author} {\bibinfo {author} {\bibfnamefont {V.}~\bibnamefont
  {Baibhav}}, \bibinfo {author} {\bibfnamefont {E.}~\bibnamefont {Berti}},
  \bibinfo {author} {\bibfnamefont {V.}~\bibnamefont {Cardoso}}, \ and\
  \bibinfo {author} {\bibfnamefont {G.}~\bibnamefont {Khanna}},\ }\href
  {\doibase 10.1103/PhysRevD.97.044048} {\bibfield  {journal} {\bibinfo
  {journal} {Phys. Rev. D}\ }\textbf {\bibinfo {volume} {97}},\ \bibinfo
  {pages} {044048} (\bibinfo {year} {2018})},\ \Eprint
  {http://arxiv.org/abs/1710.02156} {arXiv:1710.02156 [gr-qc]} \BibitemShut
  {NoStop}%
\bibitem [{\citenamefont {London}\ \emph {et~al.}(2014)\citenamefont {London},
  \citenamefont {Shoemaker},\ and\ \citenamefont {Healy}}]{London:2014cma}%
  \BibitemOpen
  \bibfield  {author} {\bibinfo {author} {\bibfnamefont {L.}~\bibnamefont
  {London}}, \bibinfo {author} {\bibfnamefont {D.}~\bibnamefont {Shoemaker}}, \
  and\ \bibinfo {author} {\bibfnamefont {J.}~\bibnamefont {Healy}},\ }\href
  {\doibase 10.1103/PhysRevD.90.124032} {\bibfield  {journal} {\bibinfo
  {journal} {Phys. Rev. D}\ }\textbf {\bibinfo {volume} {90}},\ \bibinfo
  {pages} {124032} (\bibinfo {year} {2014})},\ \bibinfo {note} {[Erratum:
  Phys.Rev.D 94, 069902 (2016)]},\ \Eprint {http://arxiv.org/abs/1404.3197}
  {arXiv:1404.3197 [gr-qc]} \BibitemShut {NoStop}%
\bibitem [{\citenamefont {Giesler}\ \emph {et~al.}(2019)\citenamefont
  {Giesler}, \citenamefont {Isi}, \citenamefont {Scheel},\ and\ \citenamefont
  {Teukolsky}}]{Giesler:2019uxc}%
  \BibitemOpen
  \bibfield  {author} {\bibinfo {author} {\bibfnamefont {M.}~\bibnamefont
  {Giesler}}, \bibinfo {author} {\bibfnamefont {M.}~\bibnamefont {Isi}},
  \bibinfo {author} {\bibfnamefont {M.~A.}\ \bibnamefont {Scheel}}, \ and\
  \bibinfo {author} {\bibfnamefont {S.}~\bibnamefont {Teukolsky}},\ }\href
  {\doibase 10.1103/PhysRevX.9.041060} {\bibfield  {journal} {\bibinfo
  {journal} {Phys. Rev.}\ }\textbf {\bibinfo {volume} {X9}},\ \bibinfo {pages}
  {041060} (\bibinfo {year} {2019})},\ \Eprint
  {http://arxiv.org/abs/1903.08284} {arXiv:1903.08284 [gr-qc]} \BibitemShut
  {NoStop}%
\bibitem [{\citenamefont {Bhagwat}\ \emph {et~al.}(2020)\citenamefont
  {Bhagwat}, \citenamefont {Forteza}, \citenamefont {Pani},\ and\ \citenamefont
  {Ferrari}}]{Bhagwat:2019dtm}%
  \BibitemOpen
  \bibfield  {author} {\bibinfo {author} {\bibfnamefont {S.}~\bibnamefont
  {Bhagwat}}, \bibinfo {author} {\bibfnamefont {X.~J.}\ \bibnamefont
  {Forteza}}, \bibinfo {author} {\bibfnamefont {P.}~\bibnamefont {Pani}}, \
  and\ \bibinfo {author} {\bibfnamefont {V.}~\bibnamefont {Ferrari}},\ }\href
  {\doibase 10.1103/PhysRevD.101.044033} {\bibfield  {journal} {\bibinfo
  {journal} {Phys. Rev. D}\ }\textbf {\bibinfo {volume} {101}},\ \bibinfo
  {pages} {044033} (\bibinfo {year} {2020})},\ \Eprint
  {http://arxiv.org/abs/1910.08708} {arXiv:1910.08708 [gr-qc]} \BibitemShut
  {NoStop}%
\bibitem [{\citenamefont {Jim\'enez~Forteza}\ \emph {et~al.}(2020)\citenamefont
  {Jim\'enez~Forteza}, \citenamefont {Bhagwat}, \citenamefont {Pani},\ and\
  \citenamefont {Ferrari}}]{JimenezForteza:2020cve}%
  \BibitemOpen
  \bibfield  {author} {\bibinfo {author} {\bibfnamefont {X.}~\bibnamefont
  {Jim\'enez~Forteza}}, \bibinfo {author} {\bibfnamefont {S.}~\bibnamefont
  {Bhagwat}}, \bibinfo {author} {\bibfnamefont {P.}~\bibnamefont {Pani}}, \
  and\ \bibinfo {author} {\bibfnamefont {V.}~\bibnamefont {Ferrari}},\ }\href
  {\doibase 10.1103/PhysRevD.102.044053} {\bibfield  {journal} {\bibinfo
  {journal} {Phys. Rev. D}\ }\textbf {\bibinfo {volume} {102}},\ \bibinfo
  {pages} {044053} (\bibinfo {year} {2020})},\ \Eprint
  {http://arxiv.org/abs/2005.03260} {arXiv:2005.03260 [gr-qc]} \BibitemShut
  {NoStop}%
\bibitem [{\citenamefont {Bustillo}\ \emph {et~al.}(2021)\citenamefont
  {Bustillo}, \citenamefont {Lasky},\ and\ \citenamefont
  {Thrane}}]{Bustillo:2020buq}%
  \BibitemOpen
  \bibfield  {author} {\bibinfo {author} {\bibfnamefont {J.~C.}\ \bibnamefont
  {Bustillo}}, \bibinfo {author} {\bibfnamefont {P.~D.}\ \bibnamefont {Lasky}},
  \ and\ \bibinfo {author} {\bibfnamefont {E.}~\bibnamefont {Thrane}},\ }\href
  {\doibase 10.1103/PhysRevD.103.024041} {\bibfield  {journal} {\bibinfo
  {journal} {Phys. Rev. D}\ }\textbf {\bibinfo {volume} {103}},\ \bibinfo
  {pages} {024041} (\bibinfo {year} {2021})},\ \Eprint
  {http://arxiv.org/abs/2010.01857} {arXiv:2010.01857 [gr-qc]} \BibitemShut
  {NoStop}%
\bibitem [{\citenamefont {Okounkova}(2020)}]{Okounkova:2020vwu}%
  \BibitemOpen
  \bibfield  {author} {\bibinfo {author} {\bibfnamefont {M.}~\bibnamefont
  {Okounkova}},\ }\href@noop {} {\  (\bibinfo {year} {2020})},\ \Eprint
  {http://arxiv.org/abs/2004.00671} {arXiv:2004.00671 [gr-qc]} \BibitemShut
  {NoStop}%
\bibitem [{\citenamefont {Mourier}\ \emph {et~al.}(2021)\citenamefont
  {Mourier}, \citenamefont {Jim\'enez~Forteza}, \citenamefont {Pook-Kolb},
  \citenamefont {Krishnan},\ and\ \citenamefont {Schnetter}}]{Mourier:2020mwa}%
  \BibitemOpen
  \bibfield  {author} {\bibinfo {author} {\bibfnamefont {P.}~\bibnamefont
  {Mourier}}, \bibinfo {author} {\bibfnamefont {X.}~\bibnamefont
  {Jim\'enez~Forteza}}, \bibinfo {author} {\bibfnamefont {D.}~\bibnamefont
  {Pook-Kolb}}, \bibinfo {author} {\bibfnamefont {B.}~\bibnamefont {Krishnan}},
  \ and\ \bibinfo {author} {\bibfnamefont {E.}~\bibnamefont {Schnetter}},\
  }\href {\doibase 10.1103/PhysRevD.103.044054} {\bibfield  {journal} {\bibinfo
   {journal} {Phys. Rev. D}\ }\textbf {\bibinfo {volume} {103}},\ \bibinfo
  {pages} {044054} (\bibinfo {year} {2021})},\ \Eprint
  {http://arxiv.org/abs/2010.15186} {arXiv:2010.15186 [gr-qc]} \BibitemShut
  {NoStop}%
\bibitem [{\citenamefont {Cook}(2020)}]{Cook:2020otn}%
  \BibitemOpen
  \bibfield  {author} {\bibinfo {author} {\bibfnamefont {G.~B.}\ \bibnamefont
  {Cook}},\ }\href {\doibase 10.1103/PhysRevD.102.024027} {\bibfield  {journal}
  {\bibinfo  {journal} {Phys. Rev. D}\ }\textbf {\bibinfo {volume} {102}},\
  \bibinfo {pages} {024027} (\bibinfo {year} {2020})},\ \Eprint
  {http://arxiv.org/abs/2004.08347} {arXiv:2004.08347 [gr-qc]} \BibitemShut
  {NoStop}%
\bibitem [{\citenamefont {Dhani}(2021)}]{Dhani:2020nik}%
  \BibitemOpen
  \bibfield  {author} {\bibinfo {author} {\bibfnamefont {A.}~\bibnamefont
  {Dhani}},\ }\href {\doibase 10.1103/PhysRevD.103.104048} {\bibfield
  {journal} {\bibinfo  {journal} {Phys. Rev. D}\ }\textbf {\bibinfo {volume}
  {103}},\ \bibinfo {pages} {104048} (\bibinfo {year} {2021})},\ \Eprint
  {http://arxiv.org/abs/2010.08602} {arXiv:2010.08602 [gr-qc]} \BibitemShut
  {NoStop}%
\bibitem [{\citenamefont {Dhani}\ and\ \citenamefont
  {Sathyaprakash}(2021)}]{Dhani:2021vac}%
  \BibitemOpen
  \bibfield  {author} {\bibinfo {author} {\bibfnamefont {A.}~\bibnamefont
  {Dhani}}\ and\ \bibinfo {author} {\bibfnamefont {B.~S.}\ \bibnamefont
  {Sathyaprakash}},\ }\href@noop {} {\  (\bibinfo {year} {2021})},\ \Eprint
  {http://arxiv.org/abs/2107.14195} {arXiv:2107.14195 [gr-qc]} \BibitemShut
  {NoStop}%
\bibitem [{\citenamefont {Finch}\ and\ \citenamefont
  {Moore}(2021)}]{Finch:2021iip}%
  \BibitemOpen
  \bibfield  {author} {\bibinfo {author} {\bibfnamefont {E.}~\bibnamefont
  {Finch}}\ and\ \bibinfo {author} {\bibfnamefont {C.~J.}\ \bibnamefont
  {Moore}},\ }\href {\doibase 10.1103/PhysRevD.103.084048} {\bibfield
  {journal} {\bibinfo  {journal} {Phys. Rev. D}\ }\textbf {\bibinfo {volume}
  {103}},\ \bibinfo {pages} {084048} (\bibinfo {year} {2021})},\ \Eprint
  {http://arxiv.org/abs/2102.07794} {arXiv:2102.07794 [gr-qc]} \BibitemShut
  {NoStop}%
\bibitem [{\citenamefont {Maga\~na Zertuche}\ \emph {et~al.}(2021)\citenamefont
  {Maga\~na Zertuche} \emph {et~al.}}]{Zertuche:2021xkb}%
  \BibitemOpen
  \bibfield  {author} {\bibinfo {author} {\bibfnamefont {L.}~\bibnamefont
  {Maga\~na Zertuche}} \emph {et~al.},\ }\href@noop {} {\  (\bibinfo {year}
  {2021})},\ \Eprint {http://arxiv.org/abs/2110.15922} {arXiv:2110.15922
  [gr-qc]} \BibitemShut {NoStop}%
\bibitem [{\citenamefont {Isi}\ \emph {et~al.}(2019)\citenamefont {Isi},
  \citenamefont {Giesler}, \citenamefont {Farr}, \citenamefont {Scheel},\ and\
  \citenamefont {Teukolsky}}]{Isi:2019aib}%
  \BibitemOpen
  \bibfield  {author} {\bibinfo {author} {\bibfnamefont {M.}~\bibnamefont
  {Isi}}, \bibinfo {author} {\bibfnamefont {M.}~\bibnamefont {Giesler}},
  \bibinfo {author} {\bibfnamefont {W.~M.}\ \bibnamefont {Farr}}, \bibinfo
  {author} {\bibfnamefont {M.~A.}\ \bibnamefont {Scheel}}, \ and\ \bibinfo
  {author} {\bibfnamefont {S.~A.}\ \bibnamefont {Teukolsky}},\ }\href {\doibase
  10.1103/PhysRevLett.123.111102} {\bibfield  {journal} {\bibinfo  {journal}
  {Phys. Rev. Lett.}\ }\textbf {\bibinfo {volume} {123}},\ \bibinfo {pages}
  {111102} (\bibinfo {year} {2019})},\ \Eprint
  {http://arxiv.org/abs/1905.00869} {arXiv:1905.00869 [gr-qc]} \BibitemShut
  {NoStop}%
\bibitem [{\citenamefont {Carullo}\ \emph {et~al.}(2019)\citenamefont
  {Carullo}, \citenamefont {Del~Pozzo},\ and\ \citenamefont
  {Veitch}}]{Carullo:2019flw}%
  \BibitemOpen
  \bibfield  {author} {\bibinfo {author} {\bibfnamefont {G.}~\bibnamefont
  {Carullo}}, \bibinfo {author} {\bibfnamefont {W.}~\bibnamefont {Del~Pozzo}},
  \ and\ \bibinfo {author} {\bibfnamefont {J.}~\bibnamefont {Veitch}},\ }\href
  {\doibase 10.1103/PhysRevD.99.123029} {\bibfield  {journal} {\bibinfo
  {journal} {Phys. Rev. D}\ }\textbf {\bibinfo {volume} {99}},\ \bibinfo
  {pages} {123029} (\bibinfo {year} {2019})},\ \bibinfo {note} {[Erratum:
  Phys.Rev.D 100, 089903 (2019)]},\ \Eprint {http://arxiv.org/abs/1902.07527}
  {arXiv:1902.07527 [gr-qc]} \BibitemShut {NoStop}%
\bibitem [{\citenamefont {Li}\ \emph {et~al.}(2021)\citenamefont {Li},
  \citenamefont {Sun}, \citenamefont {Lo}, \citenamefont {Payne},\ and\
  \citenamefont {Chen}}]{Li:2021wgz}%
  \BibitemOpen
  \bibfield  {author} {\bibinfo {author} {\bibfnamefont {X.}~\bibnamefont
  {Li}}, \bibinfo {author} {\bibfnamefont {L.}~\bibnamefont {Sun}}, \bibinfo
  {author} {\bibfnamefont {R.~K.~L.}\ \bibnamefont {Lo}}, \bibinfo {author}
  {\bibfnamefont {E.}~\bibnamefont {Payne}}, \ and\ \bibinfo {author}
  {\bibfnamefont {Y.}~\bibnamefont {Chen}},\ }\href@noop {} {\  (\bibinfo
  {year} {2021})},\ \Eprint {http://arxiv.org/abs/2110.03116} {arXiv:2110.03116
  [gr-qc]} \BibitemShut {NoStop}%
\bibitem [{\citenamefont {Maggiore}(2007)}]{Maggiore:2007ulw}%
  \BibitemOpen
  \bibfield  {author} {\bibinfo {author} {\bibfnamefont {M.}~\bibnamefont
  {Maggiore}},\ }\href@noop {} {\emph {\bibinfo {title} {{Gravitational Waves.
  Vol. 1: Theory and Experiments}}}},\ Oxford Master Series in Physics\
  (\bibinfo  {publisher} {Oxford University Press},\ \bibinfo {year}
  {2007})\BibitemShut {NoStop}%
\bibitem [{\citenamefont {Carullo}\ \emph {et~al.}()\citenamefont {Carullo},
  \citenamefont {Pozzo}, \citenamefont {Laghi}, \citenamefont {Isi},\ and\
  \citenamefont {Veitch}}]{pyRing}%
  \BibitemOpen
  \bibfield  {author} {\bibinfo {author} {\bibfnamefont {G.}~\bibnamefont
  {Carullo}}, \bibinfo {author} {\bibfnamefont {W.~D.}\ \bibnamefont {Pozzo}},
  \bibinfo {author} {\bibfnamefont {D.}~\bibnamefont {Laghi}}, \bibinfo
  {author} {\bibfnamefont {M.}~\bibnamefont {Isi}}, \ and\ \bibinfo {author}
  {\bibfnamefont {J.}~\bibnamefont {Veitch}},\ }\href@noop {} {\enquote
  {\bibinfo {title} {\texttt{pyRing}: a time-domain ringdown analysis python
  package},}\ }\bibinfo {howpublished}
  {\url{https://git.ligo.org/lscsoft/pyring}}\BibitemShut {NoStop}%
\bibitem [{\citenamefont {Del~Pozzo}\ and\ \citenamefont {Veitch}()}]{cpnest}%
  \BibitemOpen
  \bibfield  {author} {\bibinfo {author} {\bibfnamefont {W.}~\bibnamefont
  {Del~Pozzo}}\ and\ \bibinfo {author} {\bibfnamefont {J.}~\bibnamefont
  {Veitch}},\ }\href@noop {} {\enquote {\bibinfo {title} {{\texttt{CPNest}: an
  efficient python parallelizable nested sampling algorithm}},}\ }\bibinfo
  {howpublished} {\url{https://github.com/johnveitch/cpnest}}\BibitemShut
  {NoStop}%
\bibitem [{\citenamefont {Isi}\ and\ \citenamefont {Farr}(2021)}]{Isi:2021iql}%
  \BibitemOpen
  \bibfield  {author} {\bibinfo {author} {\bibfnamefont {M.}~\bibnamefont
  {Isi}}\ and\ \bibinfo {author} {\bibfnamefont {W.~M.}\ \bibnamefont {Farr}},\
  }\href@noop {} {\  (\bibinfo {year} {2021})},\ \Eprint
  {http://arxiv.org/abs/2107.05609} {arXiv:2107.05609 [gr-qc]} \BibitemShut
  {NoStop}%
\bibitem [{\citenamefont {Abbott}\ \emph
  {et~al.}(2021{\natexlab{g}})\citenamefont {Abbott} \emph
  {et~al.}}]{LIGOScientific:2019lzm}%
  \BibitemOpen
  \bibfield  {author} {\bibinfo {author} {\bibfnamefont {R.}~\bibnamefont
  {Abbott}} \emph {et~al.} (\bibinfo {collaboration} {LIGO Scientific,
  Virgo}),\ }\href {\doibase 10.1016/j.softx.2021.100658} {\bibfield  {journal}
  {\bibinfo  {journal} {SoftwareX}\ }\textbf {\bibinfo {volume} {13}},\
  \bibinfo {pages} {100658} (\bibinfo {year} {2021}{\natexlab{g}})},\ \Eprint
  {http://arxiv.org/abs/1912.11716} {arXiv:1912.11716 [gr-qc]} \BibitemShut
  {NoStop}%
\bibitem [{\citenamefont {Cotesta}\ \emph {et~al.}()\citenamefont {Cotesta},
  \citenamefont {Carullo}, \citenamefont {Berti},\ and\ \citenamefont
  {Cardoso}}]{inprep}%
  \BibitemOpen
  \bibfield  {author} {\bibinfo {author} {\bibfnamefont {R.}~\bibnamefont
  {Cotesta}}, \bibinfo {author} {\bibfnamefont {G.}~\bibnamefont {Carullo}},
  \bibinfo {author} {\bibfnamefont {E.}~\bibnamefont {Berti}}, \ and\ \bibinfo
  {author} {\bibfnamefont {V.}~\bibnamefont {Cardoso}},\ }\href@noop {}
  {}\bibinfo {howpublished} {in preparation}\BibitemShut {NoStop}%
\bibitem [{O1_()}]{O1_samples_release}%
  \BibitemOpen
  \href@noop {} {}\bibinfo {howpublished}
  {\url{https://dcc.ligo.org/LIGO-P1800370/public}}\BibitemShut {NoStop}%
\bibitem [{\citenamefont {Bohé}\ \emph {et~al.}(2017)\citenamefont {Bohé}
  \emph {et~al.}}]{Bohe:2016gbl}%
  \BibitemOpen
  \bibfield  {author} {\bibinfo {author} {\bibfnamefont {A.}~\bibnamefont
  {Bohé}} \emph {et~al.},\ }\href {\doibase 10.1103/PhysRevD.95.044028}
  {\bibfield  {journal} {\bibinfo  {journal} {Phys. Rev.}\ }\textbf {\bibinfo
  {volume} {D95}},\ \bibinfo {pages} {044028} (\bibinfo {year} {2017})},\
  \Eprint {http://arxiv.org/abs/1611.03703} {arXiv:1611.03703 [gr-qc]}
  \BibitemShut {NoStop}%
\bibitem [{\citenamefont {Husa}\ \emph {et~al.}(2016)\citenamefont {Husa},
  \citenamefont {Khan}, \citenamefont {Hannam}, \citenamefont {P\"urrer},
  \citenamefont {Ohme}, \citenamefont {Jim\'enez~Forteza},\ and\ \citenamefont
  {Boh\'e}}]{Husa:2015iqa}%
  \BibitemOpen
  \bibfield  {author} {\bibinfo {author} {\bibfnamefont {S.}~\bibnamefont
  {Husa}}, \bibinfo {author} {\bibfnamefont {S.}~\bibnamefont {Khan}}, \bibinfo
  {author} {\bibfnamefont {M.}~\bibnamefont {Hannam}}, \bibinfo {author}
  {\bibfnamefont {M.}~\bibnamefont {P\"urrer}}, \bibinfo {author}
  {\bibfnamefont {F.}~\bibnamefont {Ohme}}, \bibinfo {author} {\bibfnamefont
  {X.}~\bibnamefont {Jim\'enez~Forteza}}, \ and\ \bibinfo {author}
  {\bibfnamefont {A.}~\bibnamefont {Boh\'e}},\ }\href {\doibase
  10.1103/PhysRevD.93.044006} {\bibfield  {journal} {\bibinfo  {journal} {Phys.
  Rev. D}\ }\textbf {\bibinfo {volume} {93}},\ \bibinfo {pages} {044006}
  (\bibinfo {year} {2016})},\ \Eprint {http://arxiv.org/abs/1508.07250}
  {arXiv:1508.07250 [gr-qc]} \BibitemShut {NoStop}%
\bibitem [{\citenamefont {Khan}\ \emph {et~al.}(2016)\citenamefont {Khan},
  \citenamefont {Husa}, \citenamefont {Hannam}, \citenamefont {Ohme},
  \citenamefont {P\"urrer}, \citenamefont {Jim\'enez~Forteza},\ and\
  \citenamefont {Boh\'e}}]{Khan:2015jqa}%
  \BibitemOpen
  \bibfield  {author} {\bibinfo {author} {\bibfnamefont {S.}~\bibnamefont
  {Khan}}, \bibinfo {author} {\bibfnamefont {S.}~\bibnamefont {Husa}}, \bibinfo
  {author} {\bibfnamefont {M.}~\bibnamefont {Hannam}}, \bibinfo {author}
  {\bibfnamefont {F.}~\bibnamefont {Ohme}}, \bibinfo {author} {\bibfnamefont
  {M.}~\bibnamefont {P\"urrer}}, \bibinfo {author} {\bibfnamefont
  {X.}~\bibnamefont {Jim\'enez~Forteza}}, \ and\ \bibinfo {author}
  {\bibfnamefont {A.}~\bibnamefont {Boh\'e}},\ }\href {\doibase
  10.1103/PhysRevD.93.044007} {\bibfield  {journal} {\bibinfo  {journal} {Phys.
  Rev. D}\ }\textbf {\bibinfo {volume} {93}},\ \bibinfo {pages} {044007}
  (\bibinfo {year} {2016})},\ \Eprint {http://arxiv.org/abs/1508.07253}
  {arXiv:1508.07253 [gr-qc]} \BibitemShut {NoStop}%
\bibitem [{\citenamefont {Hannam}\ \emph {et~al.}(2014)\citenamefont {Hannam},
  \citenamefont {Schmidt}, \citenamefont {Boh\'e}, \citenamefont {Haegel},
  \citenamefont {Husa}, \citenamefont {Ohme}, \citenamefont {Pratten},\ and\
  \citenamefont {P\"urrer}}]{Hannam:2013oca}%
  \BibitemOpen
  \bibfield  {author} {\bibinfo {author} {\bibfnamefont {M.}~\bibnamefont
  {Hannam}}, \bibinfo {author} {\bibfnamefont {P.}~\bibnamefont {Schmidt}},
  \bibinfo {author} {\bibfnamefont {A.}~\bibnamefont {Boh\'e}}, \bibinfo
  {author} {\bibfnamefont {L.}~\bibnamefont {Haegel}}, \bibinfo {author}
  {\bibfnamefont {S.}~\bibnamefont {Husa}}, \bibinfo {author} {\bibfnamefont
  {F.}~\bibnamefont {Ohme}}, \bibinfo {author} {\bibfnamefont {G.}~\bibnamefont
  {Pratten}}, \ and\ \bibinfo {author} {\bibfnamefont {M.}~\bibnamefont
  {P\"urrer}},\ }\href {\doibase 10.1103/PhysRevLett.113.151101} {\bibfield
  {journal} {\bibinfo  {journal} {Phys. Rev. Lett.}\ }\textbf {\bibinfo
  {volume} {113}},\ \bibinfo {pages} {151101} (\bibinfo {year} {2014})},\
  \Eprint {http://arxiv.org/abs/1308.3271} {arXiv:1308.3271 [gr-qc]}
  \BibitemShut {NoStop}%
\bibitem [{\citenamefont {Boyle}\ \emph {et~al.}(2019)\citenamefont {Boyle}
  \emph {et~al.}}]{Boyle:2019kee}%
  \BibitemOpen
  \bibfield  {author} {\bibinfo {author} {\bibfnamefont {M.}~\bibnamefont
  {Boyle}} \emph {et~al.},\ }\href {\doibase 10.1088/1361-6382/ab34e2}
  {\bibfield  {journal} {\bibinfo  {journal} {Class. Quant. Grav.}\ }\textbf
  {\bibinfo {volume} {36}},\ \bibinfo {pages} {195006} (\bibinfo {year}
  {2019})},\ \Eprint {http://arxiv.org/abs/1904.04831} {arXiv:1904.04831
  [gr-qc]} \BibitemShut {NoStop}%
\bibitem [{\citenamefont {Sberna}\ \emph {et~al.}(2021)\citenamefont {Sberna},
  \citenamefont {Bosch}, \citenamefont {East}, \citenamefont {Green},\ and\
  \citenamefont {Lehner}}]{Sberna:2021eui}%
  \BibitemOpen
  \bibfield  {author} {\bibinfo {author} {\bibfnamefont {L.}~\bibnamefont
  {Sberna}}, \bibinfo {author} {\bibfnamefont {P.}~\bibnamefont {Bosch}},
  \bibinfo {author} {\bibfnamefont {W.~E.}\ \bibnamefont {East}}, \bibinfo
  {author} {\bibfnamefont {S.~R.}\ \bibnamefont {Green}}, \ and\ \bibinfo
  {author} {\bibfnamefont {L.}~\bibnamefont {Lehner}},\ }\href@noop {} {\
  (\bibinfo {year} {2021})},\ \Eprint {http://arxiv.org/abs/2112.11168}
  {arXiv:2112.11168 [gr-qc]} \BibitemShut {NoStop}%
\bibitem [{\citenamefont {Bhagwat}\ \emph {et~al.}(2018)\citenamefont
  {Bhagwat}, \citenamefont {Okounkova}, \citenamefont {Ballmer}, \citenamefont
  {Brown}, \citenamefont {Giesler}, \citenamefont {Scheel},\ and\ \citenamefont
  {Teukolsky}}]{Bhagwat:2017tkm}%
  \BibitemOpen
  \bibfield  {author} {\bibinfo {author} {\bibfnamefont {S.}~\bibnamefont
  {Bhagwat}}, \bibinfo {author} {\bibfnamefont {M.}~\bibnamefont {Okounkova}},
  \bibinfo {author} {\bibfnamefont {S.~W.}\ \bibnamefont {Ballmer}}, \bibinfo
  {author} {\bibfnamefont {D.~A.}\ \bibnamefont {Brown}}, \bibinfo {author}
  {\bibfnamefont {M.}~\bibnamefont {Giesler}}, \bibinfo {author} {\bibfnamefont
  {M.~A.}\ \bibnamefont {Scheel}}, \ and\ \bibinfo {author} {\bibfnamefont
  {S.~A.}\ \bibnamefont {Teukolsky}},\ }\href {\doibase
  10.1103/PhysRevD.97.104065} {\bibfield  {journal} {\bibinfo  {journal} {Phys.
  Rev. D}\ }\textbf {\bibinfo {volume} {97}},\ \bibinfo {pages} {104065}
  (\bibinfo {year} {2018})},\ \Eprint {http://arxiv.org/abs/1711.00926}
  {arXiv:1711.00926 [gr-qc]} \BibitemShut {NoStop}%
\bibitem [{\citenamefont {Dorband}\ \emph {et~al.}(2006)\citenamefont
  {Dorband}, \citenamefont {Berti}, \citenamefont {Diener}, \citenamefont
  {Schnetter},\ and\ \citenamefont {Tiglio}}]{Dorband:2006gg}%
  \BibitemOpen
  \bibfield  {author} {\bibinfo {author} {\bibfnamefont {E.~N.}\ \bibnamefont
  {Dorband}}, \bibinfo {author} {\bibfnamefont {E.}~\bibnamefont {Berti}},
  \bibinfo {author} {\bibfnamefont {P.}~\bibnamefont {Diener}}, \bibinfo
  {author} {\bibfnamefont {E.}~\bibnamefont {Schnetter}}, \ and\ \bibinfo
  {author} {\bibfnamefont {M.}~\bibnamefont {Tiglio}},\ }\href {\doibase
  10.1103/PhysRevD.74.084028} {\bibfield  {journal} {\bibinfo  {journal} {Phys.
  Rev. D}\ }\textbf {\bibinfo {volume} {74}},\ \bibinfo {pages} {084028}
  (\bibinfo {year} {2006})},\ \Eprint {http://arxiv.org/abs/gr-qc/0608091}
  {arXiv:gr-qc/0608091} \BibitemShut {NoStop}%
\bibitem [{\citenamefont {London}(2020)}]{London:2018gaq}%
  \BibitemOpen
  \bibfield  {author} {\bibinfo {author} {\bibfnamefont {L.~T.}\ \bibnamefont
  {London}},\ }\href {\doibase 10.1103/PhysRevD.102.084052} {\bibfield
  {journal} {\bibinfo  {journal} {Phys. Rev. D}\ }\textbf {\bibinfo {volume}
  {102}},\ \bibinfo {pages} {084052} (\bibinfo {year} {2020})},\ \Eprint
  {http://arxiv.org/abs/1801.08208} {arXiv:1801.08208 [gr-qc]} \BibitemShut
  {NoStop}%
\bibitem [{\citenamefont {Forteza}\ and\ \citenamefont
  {Mourier}(2021)}]{Forteza:2021wfq}%
  \BibitemOpen
  \bibfield  {author} {\bibinfo {author} {\bibfnamefont {X.~J.}\ \bibnamefont
  {Forteza}}\ and\ \bibinfo {author} {\bibfnamefont {P.}~\bibnamefont
  {Mourier}},\ }\href {\doibase 10.1103/PhysRevD.104.124072} {\bibfield
  {journal} {\bibinfo  {journal} {Phys. Rev. D}\ }\textbf {\bibinfo {volume}
  {104}},\ \bibinfo {pages} {124072} (\bibinfo {year} {2021})},\ \Eprint
  {http://arxiv.org/abs/2107.11829} {arXiv:2107.11829 [gr-qc]} \BibitemShut
  {NoStop}%
\bibitem [{\citenamefont {Pan}\ \emph {et~al.}(2011)\citenamefont {Pan},
  \citenamefont {Buonanno}, \citenamefont {Boyle}, \citenamefont {Buchman},
  \citenamefont {Kidder}, \citenamefont {Pfeiffer},\ and\ \citenamefont
  {Scheel}}]{Pan:2011gk}%
  \BibitemOpen
  \bibfield  {author} {\bibinfo {author} {\bibfnamefont {Y.}~\bibnamefont
  {Pan}}, \bibinfo {author} {\bibfnamefont {A.}~\bibnamefont {Buonanno}},
  \bibinfo {author} {\bibfnamefont {M.}~\bibnamefont {Boyle}}, \bibinfo
  {author} {\bibfnamefont {L.~T.}\ \bibnamefont {Buchman}}, \bibinfo {author}
  {\bibfnamefont {L.~E.}\ \bibnamefont {Kidder}}, \bibinfo {author}
  {\bibfnamefont {H.~P.}\ \bibnamefont {Pfeiffer}}, \ and\ \bibinfo {author}
  {\bibfnamefont {M.~A.}\ \bibnamefont {Scheel}},\ }\href {\doibase
  10.1103/PhysRevD.84.124052} {\bibfield  {journal} {\bibinfo  {journal} {Phys.
  Rev.}\ }\textbf {\bibinfo {volume} {D84}},\ \bibinfo {pages} {124052}
  (\bibinfo {year} {2011})},\ \Eprint {http://arxiv.org/abs/1106.1021}
  {arXiv:1106.1021 [gr-qc]} \BibitemShut {NoStop}%
\bibitem [{\citenamefont {Damour}\ and\ \citenamefont
  {Nagar}(2014)}]{Damour:2014yha}%
  \BibitemOpen
  \bibfield  {author} {\bibinfo {author} {\bibfnamefont {T.}~\bibnamefont
  {Damour}}\ and\ \bibinfo {author} {\bibfnamefont {A.}~\bibnamefont {Nagar}},\
  }\href {\doibase 10.1103/PhysRevD.90.024054} {\bibfield  {journal} {\bibinfo
  {journal} {Phys. Rev. D}\ }\textbf {\bibinfo {volume} {90}},\ \bibinfo
  {pages} {024054} (\bibinfo {year} {2014})},\ \Eprint
  {http://arxiv.org/abs/1406.0401} {arXiv:1406.0401 [gr-qc]} \BibitemShut
  {NoStop}%
\bibitem [{\citenamefont {Brito}\ \emph {et~al.}(2018)\citenamefont {Brito},
  \citenamefont {Buonanno},\ and\ \citenamefont {Raymond}}]{Brito:2018rfr}%
  \BibitemOpen
  \bibfield  {author} {\bibinfo {author} {\bibfnamefont {R.}~\bibnamefont
  {Brito}}, \bibinfo {author} {\bibfnamefont {A.}~\bibnamefont {Buonanno}}, \
  and\ \bibinfo {author} {\bibfnamefont {V.}~\bibnamefont {Raymond}},\ }\href
  {\doibase 10.1103/PhysRevD.98.084038} {\bibfield  {journal} {\bibinfo
  {journal} {Phys. Rev. D}\ }\textbf {\bibinfo {volume} {98}},\ \bibinfo
  {pages} {084038} (\bibinfo {year} {2018})},\ \Eprint
  {http://arxiv.org/abs/1805.00293} {arXiv:1805.00293 [gr-qc]} \BibitemShut
  {NoStop}%
\bibitem [{\citenamefont {Berti}\ \emph {et~al.}(2016)\citenamefont {Berti},
  \citenamefont {Sesana}, \citenamefont {Barausse}, \citenamefont {Cardoso},\
  and\ \citenamefont {Belczynski}}]{Berti:2016lat}%
  \BibitemOpen
  \bibfield  {author} {\bibinfo {author} {\bibfnamefont {E.}~\bibnamefont
  {Berti}}, \bibinfo {author} {\bibfnamefont {A.}~\bibnamefont {Sesana}},
  \bibinfo {author} {\bibfnamefont {E.}~\bibnamefont {Barausse}}, \bibinfo
  {author} {\bibfnamefont {V.}~\bibnamefont {Cardoso}}, \ and\ \bibinfo
  {author} {\bibfnamefont {K.}~\bibnamefont {Belczynski}},\ }\href {\doibase
  10.1103/PhysRevLett.117.101102} {\bibfield  {journal} {\bibinfo  {journal}
  {Phys. Rev. Lett.}\ }\textbf {\bibinfo {volume} {117}},\ \bibinfo {pages}
  {101102} (\bibinfo {year} {2016})},\ \Eprint
  {http://arxiv.org/abs/1605.09286} {arXiv:1605.09286 [gr-qc]} \BibitemShut
  {NoStop}%
\bibitem [{\citenamefont {Ota}\ and\ \citenamefont
  {Chirenti}(2021)}]{Ota:2021ypb}%
  \BibitemOpen
  \bibfield  {author} {\bibinfo {author} {\bibfnamefont {I.}~\bibnamefont
  {Ota}}\ and\ \bibinfo {author} {\bibfnamefont {C.}~\bibnamefont {Chirenti}},\
  }\href@noop {} {\  (\bibinfo {year} {2021})},\ \Eprint
  {http://arxiv.org/abs/2108.01774} {arXiv:2108.01774 [gr-qc]} \BibitemShut
  {NoStop}%
\bibitem [{\citenamefont {Isi}\ and\ \citenamefont {Farr}(2022)}]{Isi:2022mhy}%
  \BibitemOpen
  \bibfield  {author} {\bibinfo {author} {\bibfnamefont {M.}~\bibnamefont
  {Isi}}\ and\ \bibinfo {author} {\bibfnamefont {W.~M.}\ \bibnamefont {Farr}},\
  }\href@noop {} {\  (\bibinfo {year} {2022})},\ \Eprint
  {http://arxiv.org/abs/2202.02941} {arXiv:2202.02941 [gr-qc]} \BibitemShut
  {NoStop}%
\bibitem [{\citenamefont {Finch}\ and\ \citenamefont
  {Moore}(2022)}]{Finch:2022ynt}%
  \BibitemOpen
  \bibfield  {author} {\bibinfo {author} {\bibfnamefont {E.}~\bibnamefont
  {Finch}}\ and\ \bibinfo {author} {\bibfnamefont {C.~J.}\ \bibnamefont
  {Moore}},\ }\href@noop {} {\  (\bibinfo {year} {2022})},\ \Eprint
  {http://arxiv.org/abs/2205.07809} {arXiv:2205.07809 [gr-qc]} \BibitemShut
  {NoStop}%
\bibitem [{\citenamefont {Macleod}\ \emph {et~al.}(2021)\citenamefont {Macleod}
  \emph {et~al.}}]{gwpy}%
  \BibitemOpen
  \bibfield  {author} {\bibinfo {author} {\bibfnamefont {D.}~\bibnamefont
  {Macleod}} \emph {et~al.},\ }\href {\doibase 10.5281/zenodo.597016} {\enquote
  {\bibinfo {title} {gwpy/gwpy: 2.0.3},}\ } (\bibinfo {year}
  {2021})\BibitemShut {NoStop}%
\bibitem [{\citenamefont {{LIGO Scientific Collaboration}}(2018)}]{lalsuite}%
  \BibitemOpen
  \bibfield  {author} {\bibinfo {author} {\bibnamefont {{LIGO Scientific
  Collaboration}}},\ }\href {\doibase 10.7935/GT1W-FZ16} {\enquote {\bibinfo
  {title} {{LIGO} {A}lgorithm {L}ibrary - {LALS}uite},}\ }\bibinfo
  {howpublished} {free software (GPL)} (\bibinfo {year} {2018})\BibitemShut
  {NoStop}%
\bibitem [{\citenamefont {Foreman-Mackey}\ \emph {et~al.}(2021)\citenamefont
  {Foreman-Mackey} \emph {et~al.}}]{corner}%
  \BibitemOpen
  \bibfield  {author} {\bibinfo {author} {\bibfnamefont {D.}~\bibnamefont
  {Foreman-Mackey}} \emph {et~al.},\ }\href {\doibase 10.5281/zenodo.591491}
  {\enquote {\bibinfo {title} {dfm/corner.py: corner.py v.2.2.1},}\ } (\bibinfo
  {year} {2021})\BibitemShut {NoStop}%
\bibitem [{\citenamefont {Behnel}\ \emph {et~al.}(2011)\citenamefont {Behnel},
  \citenamefont {Bradshaw}, \citenamefont {Citro}, \citenamefont {Dalcin},
  \citenamefont {Seljebotn},\ and\ \citenamefont {Smith}}]{cython}%
  \BibitemOpen
  \bibfield  {author} {\bibinfo {author} {\bibfnamefont {S.}~\bibnamefont
  {Behnel}}, \bibinfo {author} {\bibfnamefont {R.}~\bibnamefont {Bradshaw}},
  \bibinfo {author} {\bibfnamefont {C.}~\bibnamefont {Citro}}, \bibinfo
  {author} {\bibfnamefont {L.}~\bibnamefont {Dalcin}}, \bibinfo {author}
  {\bibfnamefont {D.}~\bibnamefont {Seljebotn}}, \ and\ \bibinfo {author}
  {\bibfnamefont {K.}~\bibnamefont {Smith}},\ }\href {\doibase
  10.1109/MCSE.2010.118} {\bibfield  {journal} {\bibinfo  {journal} {Comput.
  Sci. Eng.}\ }\textbf {\bibinfo {volume} {13}},\ \bibinfo {pages} {31 }
  (\bibinfo {year} {2011})}\BibitemShut {NoStop}%
\bibitem [{\citenamefont {Collette}(2013)}]{h5py}%
  \BibitemOpen
  \bibfield  {author} {\bibinfo {author} {\bibfnamefont {A.}~\bibnamefont
  {Collette}},\ }\href@noop {} {\emph {\bibinfo {title} {Python and HDF5}}}\
  (\bibinfo  {publisher} {O'Reilly},\ \bibinfo {year} {2013})\BibitemShut
  {NoStop}%
\bibitem [{\citenamefont {Hunter}(2007)}]{matplotlib}%
  \BibitemOpen
  \bibfield  {author} {\bibinfo {author} {\bibfnamefont {J.~D.}\ \bibnamefont
  {Hunter}},\ }\href {\doibase 10.1109/MCSE.2007.55} {\bibfield  {journal}
  {\bibinfo  {journal} {Comput. Sci. Eng.}\ }\textbf {\bibinfo {volume} {9}},\
  \bibinfo {pages} {90} (\bibinfo {year} {2007})}\BibitemShut {NoStop}%
\bibitem [{\citenamefont {Harris}\ \emph {et~al.}(2020)\citenamefont {Harris}
  \emph {et~al.}}]{numpy}%
  \BibitemOpen
  \bibfield  {author} {\bibinfo {author} {\bibfnamefont {C.~R.}\ \bibnamefont
  {Harris}} \emph {et~al.},\ }\href {\doibase 10.1038/s41586-020-2649-2}
  {\bibfield  {journal} {\bibinfo  {journal} {Nature (London)}\ }\textbf
  {\bibinfo {volume} {585}},\ \bibinfo {pages} {357} (\bibinfo {year}
  {2020})}\BibitemShut {NoStop}%
\bibitem [{\citenamefont {Virtanen}\ \emph {et~al.}(2020)\citenamefont
  {Virtanen}, \citenamefont {Gommers}, \citenamefont {Oliphant}, \citenamefont
  {Haberland}, \citenamefont {Reddy}, \citenamefont {Cournapeau}, \citenamefont
  {Burovski}, \citenamefont {Peterson}, \citenamefont {{Weckesser}},
  \citenamefont {{Bright}}, \citenamefont {{van der Walt}}, \citenamefont
  {{Brett}}, \citenamefont {{Wilson}}, \citenamefont {{Jarrod Millman}},
  \citenamefont {{Mayorov}}, \citenamefont {{Nelson}}, \citenamefont {{Jones}},
  \citenamefont {{Kern}}, \citenamefont {{Larson}}, \citenamefont {{Carey}},
  \citenamefont {{Polat}}, \citenamefont {{Feng}}, \citenamefont {{Moore}},
  \citenamefont {{Vand erPlas}}, \citenamefont {{Laxalde}}, \citenamefont
  {{Perktold}}, \citenamefont {{Cimrman}}, \citenamefont {{Henriksen}},
  \citenamefont {{Quintero}}, \citenamefont {{Harris}}, \citenamefont
  {{Archibald}}, \citenamefont {{Ribeiro}}, \citenamefont {{Pedregosa}},
  \citenamefont {{van Mulbregt}},\ and\ \citenamefont
  {{Contributors}}}]{scipy}%
  \BibitemOpen
  \bibfield  {author} {\bibinfo {author} {\bibfnamefont {P.}~\bibnamefont
  {Virtanen}}, \bibinfo {author} {\bibfnamefont {R.}~\bibnamefont {Gommers}},
  \bibinfo {author} {\bibfnamefont {T.~E.}\ \bibnamefont {Oliphant}}, \bibinfo
  {author} {\bibfnamefont {M.}~\bibnamefont {Haberland}}, \bibinfo {author}
  {\bibfnamefont {T.}~\bibnamefont {Reddy}}, \bibinfo {author} {\bibfnamefont
  {D.}~\bibnamefont {Cournapeau}}, \bibinfo {author} {\bibfnamefont
  {E.}~\bibnamefont {Burovski}}, \bibinfo {author} {\bibfnamefont
  {P.}~\bibnamefont {Peterson}}, \bibinfo {author} {\bibfnamefont
  {W.}~\bibnamefont {{Weckesser}}}, \bibinfo {author} {\bibfnamefont
  {J.}~\bibnamefont {{Bright}}}, \bibinfo {author} {\bibfnamefont {S.~J.}\
  \bibnamefont {{van der Walt}}}, \bibinfo {author} {\bibfnamefont
  {M.}~\bibnamefont {{Brett}}}, \bibinfo {author} {\bibfnamefont
  {J.}~\bibnamefont {{Wilson}}}, \bibinfo {author} {\bibfnamefont
  {K.}~\bibnamefont {{Jarrod Millman}}}, \bibinfo {author} {\bibfnamefont
  {N.}~\bibnamefont {{Mayorov}}}, \bibinfo {author} {\bibfnamefont {A.~R.~J.}\
  \bibnamefont {{Nelson}}}, \bibinfo {author} {\bibfnamefont {E.}~\bibnamefont
  {{Jones}}}, \bibinfo {author} {\bibfnamefont {R.}~\bibnamefont {{Kern}}},
  \bibinfo {author} {\bibfnamefont {E.}~\bibnamefont {{Larson}}}, \bibinfo
  {author} {\bibfnamefont {C.}~\bibnamefont {{Carey}}}, \bibinfo {author}
  {\bibfnamefont {l.}~\bibnamefont {{Polat}}}, \bibinfo {author} {\bibfnamefont
  {Y.}~\bibnamefont {{Feng}}}, \bibinfo {author} {\bibfnamefont {E.~W.}\
  \bibnamefont {{Moore}}}, \bibinfo {author} {\bibfnamefont {J.}~\bibnamefont
  {{Vand erPlas}}}, \bibinfo {author} {\bibfnamefont {D.}~\bibnamefont
  {{Laxalde}}}, \bibinfo {author} {\bibfnamefont {J.}~\bibnamefont
  {{Perktold}}}, \bibinfo {author} {\bibfnamefont {R.}~\bibnamefont
  {{Cimrman}}}, \bibinfo {author} {\bibfnamefont {I.}~\bibnamefont
  {{Henriksen}}}, \bibinfo {author} {\bibfnamefont {E.~A.}\ \bibnamefont
  {{Quintero}}}, \bibinfo {author} {\bibfnamefont {C.~R.}\ \bibnamefont
  {{Harris}}}, \bibinfo {author} {\bibfnamefont {A.~M.}\ \bibnamefont
  {{Archibald}}}, \bibinfo {author} {\bibfnamefont {A.~H.}\ \bibnamefont
  {{Ribeiro}}}, \bibinfo {author} {\bibfnamefont {F.}~\bibnamefont
  {{Pedregosa}}}, \bibinfo {author} {\bibfnamefont {P.}~\bibnamefont {{van
  Mulbregt}}}, \ and\ \bibinfo {author} {\bibfnamefont {S.~.~.}\ \bibnamefont
  {{Contributors}}},\ }\href@noop {} {\bibfield  {journal} {\bibinfo  {journal}
  {Nature Methods}\ } (\bibinfo {year} {2020})}\BibitemShut {NoStop}%
\bibitem [{\citenamefont {Waskom}\ \emph {et~al.}(2021)\citenamefont {Waskom}
  \emph {et~al.}}]{seaborn}%
  \BibitemOpen
  \bibfield  {author} {\bibinfo {author} {\bibfnamefont {M.}~\bibnamefont
  {Waskom}} \emph {et~al.},\ }\href {\doibase 10.5281/zenodo.592845} {\enquote
  {\bibinfo {title} {mwaskom/seaborn: v0.11.2 (august 2021)},}\ } (\bibinfo
  {year} {2021})\BibitemShut {NoStop}%
\end{thebibliography}%

\clearpage

\section*{SUPPLEMENTAL MATERIAL}

\noindent {\bf \em Details of the peak time reconstruction.} 
The peak time in the Hanford detector is reconstructed by generating the $h_{+}, h_{\times}$ waveform polarizations in post-processing, using the \ac{LVK} posterior samples~\cite{O1_samples_release}, and computing the maximum of $h_{+}^2 + h_{\times}^2$. In the main text we use the peak time $t_\mathrm{start}^\mathrm{H1}$ reconstructed from the SEOBNRv4 model~\cite{Bohe:2016gbl}, but to quantify waveform systematics we have repeated the calculation using also the IMRPhenomPv2 model~\cite{Husa:2015iqa, Khan:2015jqa, Hannam:2013oca}. The resulting distribution has median $\bar{t}\,_{\mathrm{peak, Pv2}}^\mathrm{H1} = 1126259462.42371$\,s and standard deviation $\sigma_{\mathrm{Pv2}} = 0.00063 \, \mathrm{s}$, i.e., it is shifted $\sim 0.5 \mathrm{ms}$ \textit{after} the time inferred from SEOBNRv4. Thus, using the reconstruction from this alternative model would reinforce our conclusions. This difference also highlights the need to properly marginalize over the peak time when evaluating the robustness of ringdown analyses.
As a conservative choice, the \pyRing package internally approximates the analysis starting time as the point on the discretized time axis immediately 
after the $t_\mathrm{start}$, specified as a float. The high sampling rate employed ensures that no issues due to this discretization arise.

\noindent {\bf \em Details of the injection study.} 
In the injection study, we use the numerical relativity simulation \texttt{SXS:BBH:0305} from the public catalog~\cite{Boyle:2019kee} of the \ac{SXS} collaboration. This simulation was set up to reproduce the GW150914 signal. The black hole binary in the numerical waveform has mass ratio $q = 1.22$ and spins aligned with the orbital angular momentum, with dimensionless magnitudes $\chi_{1} = 0.33$ and $\chi_{2} = -0.44$. For the synthetic signal, we place the system at a luminosity distance of $D_\mathrm{L} = 410$\,Mpc and we use a redshifted total mass $M = 72 M_\odot$, in agreement with the median values estimated by the \ac{LVK} collaboration~\cite{TheLIGOScientific:2016wfe}.  Finally, the simulated signal is superimposed to the real detector noise at times $[-30,-25,-20,-5,5,10,15,20,25,30]$\,s with respect to the peak time $t_\mathrm{peak, inj}^\mathrm{H1} = 1126259472.423$\,s, approximately $10$\,s after the coalescence time of GW150914.  We use the same noise ACF used in the analysis of the GW150914 event.

\begin{figure*}[thbp]
\centering
\includegraphics[width=\textwidth]{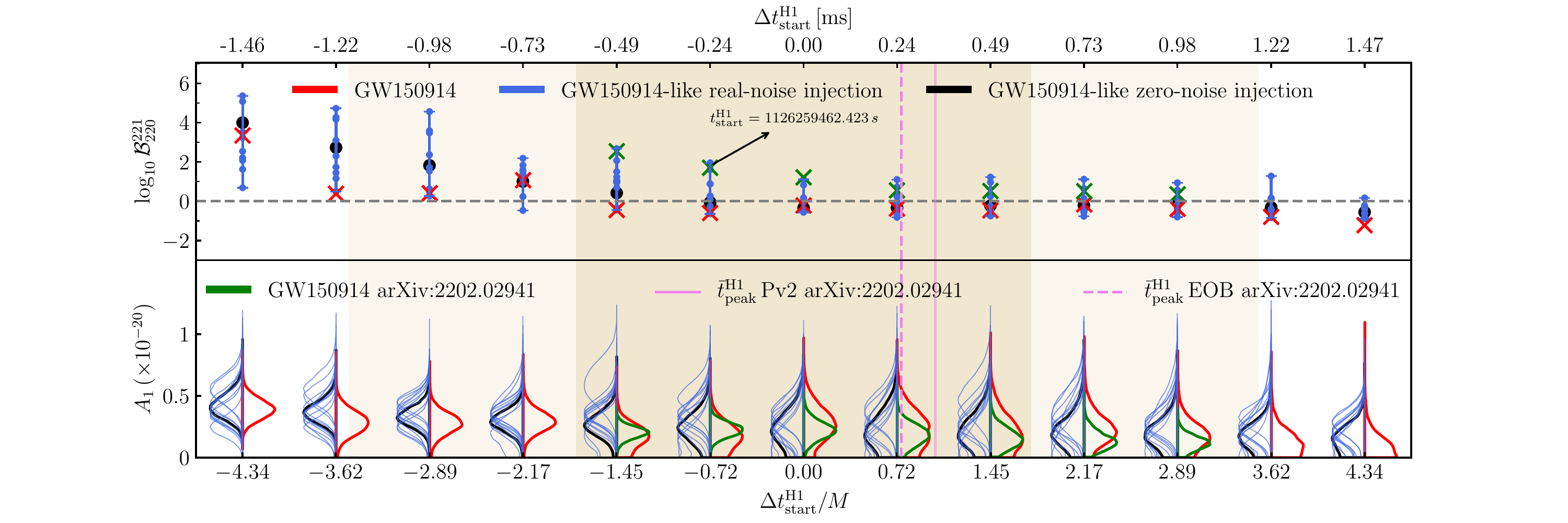}
\caption{Same as Fig.~2 in the main text, with the addition of (i) the Bayes factors estimated by Ref.~\cite{Isi:2022mhy} (green crosses in the top panel); (ii) the amplitude posteriors computed by Ref.~\cite{Isi:2022mhy} (shown in green in the bottom panel); and (iii) the estimates of $t_{\rm peak}$ by Ref.~\cite{Isi:2022mhy} obtained using either IMRPhenomPv2 (solid vertical line) or SEOBNRv4ROM (dashed vertical line).}
   \label{fig:BF_and_amp_N1_vs_N0_vary_time_new_Isi}
\end{figure*}

\begin{figure*}[thbp]
\centering
\includegraphics[width=\textwidth]{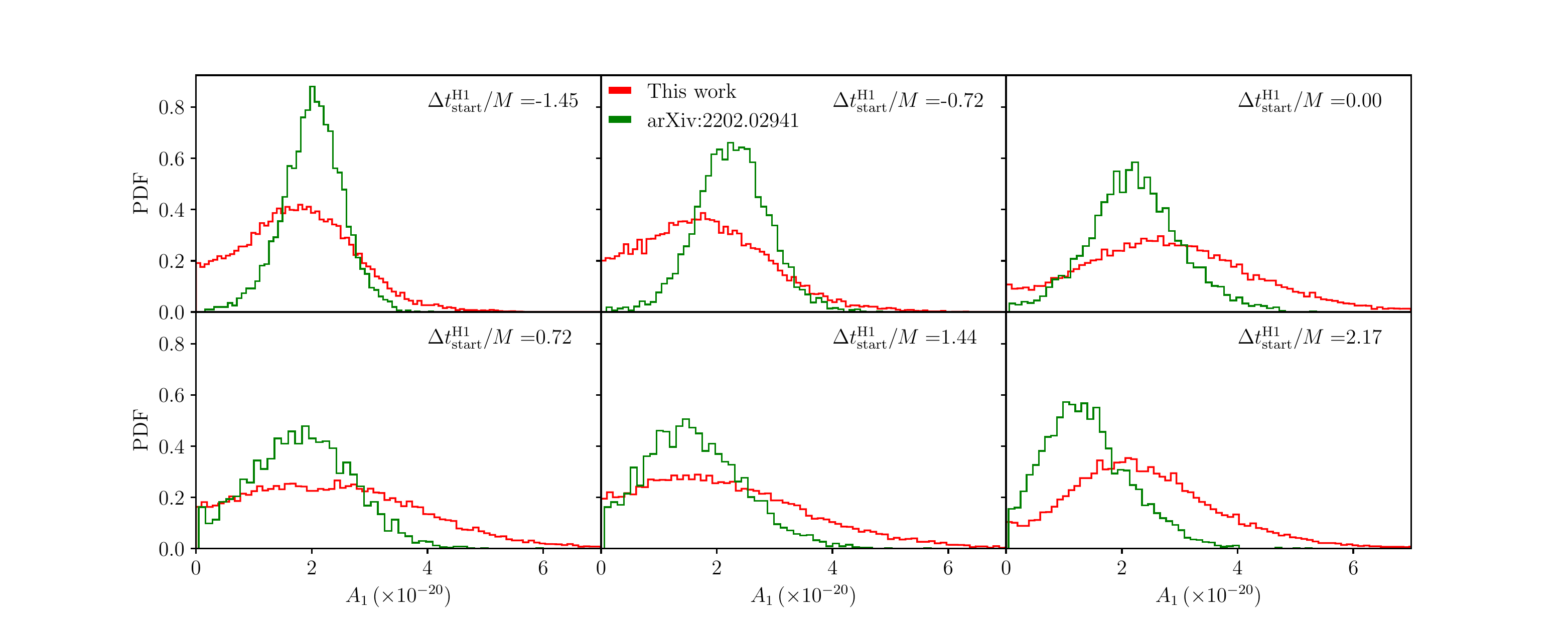}
\caption{Comparison of the posterior distributions of the overtone amplitude $A_1$ for a selected range of starting times close to the peak time estimate. Red histograms are our results (shown also in Fig.~\ref{fig:BF_and_amp_N1_vs_N0_vary_time_new_Isi}), while green histograms refer to the publicly available samples from Ref.~\cite{Isi:2022mhy}.}
   \label{fig:A1_comparison_Isi}
\end{figure*}

\noindent {\bf \em Comparison with Ref.~\cite{Isi:2022mhy}.} 
In Figs.~\ref{fig:BF_and_amp_N1_vs_N0_vary_time_new_Isi} and~\ref{fig:A1_comparison_Isi} we compare our results
to the publicly available posterior samples and Bayes factors from Ref.~\cite{Isi:2022mhy},  where the authors of~\cite{Isi:2019aib} reanalyzed GW150914.
We find their results to be broadly consistent with ours, although we disagree on the conclusions that can be drawn from these results.
The green crosses in Fig.~\ref{fig:BF_and_amp_N1_vs_N0_vary_time_new_Isi} (to be compared with Fig.~7 of~\cite{Isi:2022mhy}) show their estimates of the Bayes factors. The vertical lines in the top panel of Fig.~\ref{fig:BF_and_amp_N1_vs_N0_vary_time_new_Isi} show two different estimates of $t_{\rm peak}$ from Ref.~\cite{Isi:2022mhy}, obtained using either the IMRPhenomPv2 (solid) or SEOBNRv4ROM (dashed) waveforms: note that their own estimates of the peak time suggest that one should actually look for overtones at later times than our own estimate. The Bayes factors computed after the peak time do not significantly depart from zero in either of the two studies. 
In fact, the Bayes factors reported in Ref.~\cite{Isi:2022mhy} are always contained within the ``error bars'' determined by noise in Fig.~\ref{fig:BF_and_amp_N1_vs_N0_vary_time_new_Isi}. In conclusion, there is no robust statistical evidence for the presence of overtones.
In our computation, we have taken care to restrict the prior as much as possible (without truncating the posterior distribution), hence the objection that the Bayes factors can be made arbitrary small by enlarging the prior range does not apply to this case.

As can be seen from Fig.~\ref{fig:A1_comparison_Isi}, when allowing for uncertainties in the starting time reconstruction, the posterior distributions of the overtone amplitudes from Ref.~\cite{Isi:2022mhy} show significant railing against zero (the data used for this plot are the same data shown in Fig.~1 of~\cite{Isi:2022mhy}, which however shows smoothed distributions on a different plotting scale).
Given the statistical ($\sim 3.5 M$ at $2 \sigma$ credibility) and systematic ($\sim 1.5 M$ after the reference time of Fig.~\ref{fig:A1_comparison_Isi}, according to our analysis) uncertainties in the starting time reconstruction, the observed railing around the peak time does not allow us to conclude in favor of a confident detection of the overtone amplitude.
We also note that it is not straightforward to draw conclusions from the ratio of the median and standard deviation (see~\cite{Isi:2022mhy}) when the posterior rails against zero, as it does at late times for this event.

One difference between our results and those of Ref.~\cite{Isi:2022mhy} concerns the tails of the posteriors: our overtone amplitude posteriors have generally larger uncertainty.
Given the large number of live points we used, typically resulting in $\sim 20000$ posterior samples (compared to $\sim 2000$ of~\cite{Isi:2022mhy}), we are confident that our algorithm is correctly estimating the posterior tails.
Let us also remark that while the authors of Ref.~\cite{Isi:2022mhy} show the results of a small number of injections, they do not systematically investigate the impact of the starting time on these injections. Our analysis implies that a systematic investigation of the effect of the starting time is critical to draw reliable conclusions.

\end{document}